\pdfoutput=1
\documentclass[11pt]{article}

\usepackage[nolist]{acronym}
\usepackage{siunitx}
\usepackage{hyperref}
\usepackage{xspace}
\usepackage{lipsum}
\usepackage{graphicx}
\usepackage{enumitem}
\usepackage[ruled,vlined]{algorithm2e}
\usepackage{color}
\usepackage{tcolorbox}
\usepackage{float}
\usepackage{mathtools}
\usepackage{subcaption}
\usepackage{amsmath}
\usepackage{amsfonts}
\usepackage{amsthm}

\usepackage{stmaryrd}
\usepackage{cleveref}
\usepackage{comment}

\usepackage{algpseudocode}
\usepackage{todonotes}

\usepackage{xargs}

\usepackage{booktabs}
\usepackage{multirow}

\usepackage{booktabs} 
\usepackage{adjustbox} 
\usepackage{stfloats} 
\usepackage{comment}

\usepackage{tabularray}
\usepackage{longtable}

\usepackage{listings}

\usepackage{misc/nicodemus}

\makeatletter
\newcommand{\vast}{\bBigg@{3}}
\newcommand{\Vast}{\bBigg@{4}}
\makeatother

\crefname{assumption}{Assumption}{Assumptions}
\crefname{construction}{Construction}{Constructions}
\crefname{corollary}{Corollary}{Corollaries}
\crefname{conjecture}{Conjecture}{Conjectures}
\crefname{definition}{Definition}{Definitions}
\crefname{exmaple}{Example}{Examples}
\crefname{figure}{Figure}{Figures}
\crefname{lemma}{Lemma}{Lemmata}
\crefname{observation}{Observation}{Observations}
\crefname{proposition}{Proposition}{Propositions}
\crefname{remark}{Remark}{Remarks}
\crefname{section}{Section}{Sections}
\crefname{theorem}{Theorem}{Theorems}
\crefname{table}{Table}{Tables}
\crefname{appendix}{Appendix}{Appendices}

\newcommandx{\pcforin}[2]{\textbf{for }{#1}\textbf{ in }{#2}\colon}

\newcommand{\sig}{\sigma}
\newcommandx{\sigA}[1][1=]{\sig_{#1}}
\newcommandx{\sigS}[1][1=]{\sig_{S#1}}

\newcommandx{\model}[1][1=]{m_{#1}}

\newcommandx{\data}[1][1=]{D_{#1}}
\newcommandx{\alldata}{\mathcal{D}}
\newcommandx{\listDatasets}{\mathcal{L}_{\data}}
\newcommandx{\listHashUnlearned}{\mathcal{L}_{h}}
\newcommandx{\dataUnlearn}[1][1=]{U_{#1}}
\newcommandx{\dataUnlearnP}[1][1=]{{U_{#1}'}^{+}}
\newcommandx{\modelD}[1][1=]{\model_{\data[#1]}}

\newcommandx{\proofModel}[1][1=]{\ensuremath{\rho}_{#1}}
\newcommandx{\proofM}[1][1=]{\ensuremath{\pi}_{\model[#1]}}
\newcommandx{\proofD}[1][1=]{\ensuremath{\pi}_{\data[#1]}}

\newcommandx{\listCom}{\mathcal{L}_{\com}}
\newcommandx{\listModel}{\mathcal{L}_{\model}}
\newcommandx{\listUp}{\mathcal{L}_{\up}}

\newcommandx{\ledgerEntry}[1][1=]{L_{#1}}

\newcommandx{\up}[1][1=]{\mathsf{up}_{#1}}

\newcommandx{\hashsData}[1][1=]{\mathcal{H}_{\data[#1]}}
\newcommandx{\hashsDataP}[1][1=]{\mathcal{H}_{\data[#1]}'}
\newcommandx{\hashsDataAdd}[1][1=]{\mathcal{H}_{\dataAdd[#1]}}
\newcommandx{\hashsDataAddP}[1][1=]{\mathcal{H}_{\dataAdd[#1]}'}
\newcommandx{\hashsUnlearned}[1][1=]{\mathcal{H}_{\dataUnlearn[#1]}}
\newcommandx{\hashsUnlearnedP}[1][1=]{\mathcal{H}_{\dataUnlearn[#1]}'}
\newcommandx{\hashsUnlearnedAdd}[1][1=]{\mathcal{H}_{\dataUnlearnAdd[#1]}}
\newcommandx{\hashsUnlearnedAddP}[1][1=]{\mathcal{H}_{\dataUnlearnAdd[#1]}'}
\newcommandx{\hashsAllData}[1][1=]{\mathcal{H}_{\data[all]}}

\newcommandx{\dataAdd}[1][1=]{D^+_{#1}}
\newcommandx{\dataUnlearnAdd}[1][1=]{U^+_{#1}}
\newcommandx{\dataAddP}[1][1=]{D_{#1}^+{}'}
\newcommandx{\dataUnlearnAddP}[1][1=]{U_{#1}^+{}'}

\newcommandx{\stS}[1][1=]{\mathsf{st}_{S,#1}}
\newcommandx{\stA}[1][1=]{\mathsf{st}_{A,#1}}
\newcommandx{\stU}[1][1=]{\mathsf{st}_{u,#1}}

\newcommand{\datasample}{d}

\newcommandx{\dUnlearn}[1][1=]{\datasample_{#1}}
\newcommandx{\proofUnlearn}[1][1=]{\ensuremath{\pi}_{\dUnlearn[#1]}}

\newcommandx{\hashData}[1][1=]{h_{\data[#1]}}

\newcommandx{\hashDataUnlearn}[1][1=]{h_{\dataUnlearn[#1]}}
\newcommandx{\hashDataSample}[1][1=]{h_{\datasample_{#1}}}

\newcommandx{\hashModel}[1][1=]{h_{\modelD[#1]}}
\newcommandx{\hashLedger}[1][1=]{h_{\ledgerEntry[#1]}}

\newcommandx{\hData}[1][1=]{h_{\data[#1]}}
\newcommandx{\hModel}[1][1=]{h_{\model[#1]}}
\newcommandx{\hUnlearn}[1][1=]{h_{\dataUnlearn[#1]}}
\newcommandx{\hDataP}[1][1=]{h'_{\data[#1]}}
\newcommandx{\hModelP}[1][1=]{h'_{\model[#1]}}
\newcommandx{\hUnlearnP}[1][1=]{h'_{\dataUnlearn[#1]}}

\newcommandx{\game}[1][1=]{\mathsf{G}_{#1}}

\newtheorem{theorem}{Theorem}

\newtheorem{definition}{Definition}

\newcommandx{\com}[1][1=]{\mathsf{com}_{#1}}

\newcommandx{\comNonMember}[1][1=]{\ensuremath{\Psi}^{\scalebox{0.5}[1.0]{\( - \)}}_{#1}}

\newcommandx{\cert}[1][1=]{c_{#1}}

\newcommand{\cntIteration}{i}

\newcommandx{\noDeleteQueriesIteration}[1][1=\cntIteration]{t_{#1}}

\newcounter{ctr}

\hypersetup{
    colorlinks = false,
    hidelinks = true
}
\begin{acronym}
    \acro{MI}{\emph{Model Inversion}}
    \acro{MIA}{\emph{Membership Inference Attack}}
    \acro{SISA}{\emph{Sharded, Isolated, Sliced, and Aggregated}}
    \acro{R1CS}{\emph{Rank-1 Constraint Systems}}
    \acro{MPC}{\emph{Multi-Party Computation}}
    \acro{HE}{\emph{Homomorphic Encryption}}
    \acro{ADSs}{\emph{Authenticated Data Structures}}
    \acro{VC}{\emph{Verified Computation}}
\end{acronym}
\usepackage[final]{ACL2023}

\usepackage{times}
\usepackage{latexsym}
\usepackage[T1]{fontenc}
\usepackage[utf8]{inputenc}
\usepackage{microtype}
\usepackage{inconsolata}

\title{Bypassing LLM Watermarks with Color-Aware Substitutions}

\author{
  Qilong Wu,  Varun Chandrasekaran\\
  University of Illinois Urbana-Champaign\\
}
\begin{document}
\onecolumn
\maketitle

\pagestyle{plain}

\begin{abstract}
Watermarking approaches are proposed to identify if text being circulated is human or large language model (LLM) generated. The state-of-the-art watermarking strategy of~\citet{kirchenbauer2023watermark} biases the LLM to generate specific (``green'') tokens. However, determining the robustness of this watermarking method is an open problem. Existing attack methods fail to evade detection for longer text segments. We overcome this limitation, and propose {\em Self Color Testing-based Substitution (SCTS)}, the first ``color-aware'' attack. SCTS obtains color information by strategically prompting the watermarked LLM and comparing output tokens frequencies. It uses this information to determine token colors, and substitutes green tokens with non-green ones. In our experiments, SCTS successfully evades watermark detection using fewer number of edits than related work. Additionally, we show both theoretically and empirically that SCTS can remove the watermark for arbitrarily long watermarked text. 
\end{abstract}

\section{Introduction}
\label{sec:intro}

Large language models (LLMs) are pervasively used in applications related to education~\citep{hadi2023large} and programming~\citep{fan2023large}. As with most generative AI technologies, however, they have dual use---for generating misinformation~\citep{chen2023can} and facilitating academic dishonesty~\citep{lancaster2021academic}. Detecting LLM-generated content is a subject of immense academic interest; paraphrasing~\citet{yang2023survey}, ``detection techniques have witnessed advancements propelled by innovations in zero-shot methods, fine-tuning'' etc. resulting in various methods being proposed. 

However, most detection techniques are susceptible to a large number of false positives~\citep{aaronson2022ai}. Erroneous detection is further exacerbated when LLMs' output more closely mirrors human content~\cite{zhang2023watermarks,ghosal2023towards}. Watermarking circumvents this issue by embedding information into the generation process by manipulating the decoding step. For example, the watermark of~\citet{kirchenbauer2023watermark, kirchenbauer2023reliability} biases the logits to boost the probability of specific tokens (denoted ``green tokens'') in contrast to the remaining ``red'' tokens. The set of green tokens is determined by a hash function, and the security of this watermark is based on the hardness of finding hash collisions. Despite the emergence of more watermarking schemes~\cite{kuditipudi2023robust, christ2023undetectable}, research shows that methods that (slightly) modify logit distributions outperform others without significant output quality degradation~\cite{piet2023mark}: humans can not identify if the text is watermarked, and detection requires processing approximately one hundred tokens. Thus, for the remainder of this work, we focus on the ~\citet{kirchenbauer2023watermark} watermark and its variant by~\citet{zhao2023provable} as exemplars. 

We study {\em if watermarking approaches based on logit perturbation are robust to post-processing distortions.} This is an important question across various applications. For example, it will reliably determine if academic integrity was violated in educational settings: if the watermark is not robust, a student can easily edit the generated watermarking text, and the resulting data will fail verification. While prior works also ask this or similar questions~\citep{sadasivan2023can, lu2023large}, they do so under unreasonable settings, where edits (or post-processing distortions) to watermarked text are unconstrained (i.e., they edit large fractions of the generated text with potential degradation to output quality). They use dedicated paraphrasing models~\citep{sadasivan2023can}, or design customized prompts for post-hoc output paraphrasing~\citep{lu2023large,shi2023red}. 
Alternatively, they design prompts to generate hard-to-detect text (which is often contextually unrelated), sometimes with low entropy~\citep{lu2023large, shi2023red}.

\begin{figure}[htbp]
\centering
\includegraphics[width=0.7\linewidth]{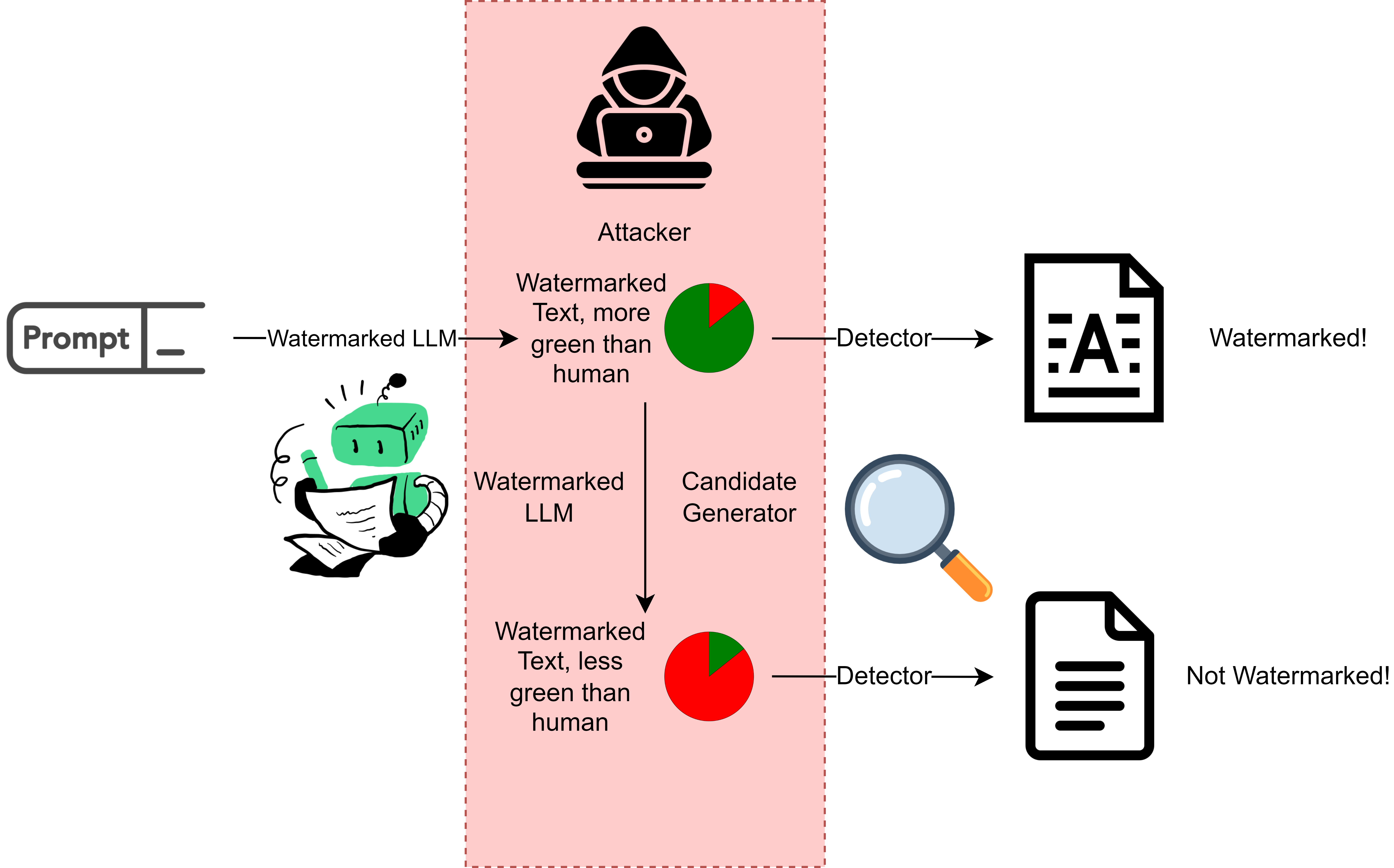}
\caption{Illustration of the setting for SCTS. The red box indicates the attacker's capability.\vspace{-5mm}}
\label{fig:intro}
\end{figure}

We evaluate robustness under a more realistic setup: post-hoc editing involves making a ``reasonable'' number of edits to maximally retain information generated by the LLM. To achieve our goal, we make two observations. First, previous attack methods are {\em color-agnostic}: existing approaches that replace text via paraphrasing still preserve some fraction of the original text, resulting in dilution rather than erasure of watermarks~\citep{kirchenbauer2023reliability} (i.e., a larger number of edits is needed for erasure). Even if they manage to evade detection for certain samples by removing ``most'' fragments, our experiments demonstrate that they would fail when constrained by a smaller edit budget (\S~\ref{section:results}). Second, prior attacks require another high-quality ``unwatermarked'' LLM for paraphrase generation which generates verbose text, and is not very practical as it can be used for generation in the first place. 

Building atop these observations, we propose the {\bf Self Color Testing-based Substitution (SCTS)} attack. This introduces new text fragments that contain fewer green tokens than human-written text, with high probability. We prompt the LLM to perform targeted (periodic) generations given an input context (refer prompt in \S~\ref{sec:our}) to get color information from the frequency of words in the generated text. This is used to perform color-aware substitution, and in turn, neutralizes the higher number of green list tokens from the preserved fragments. Consequently, SCTS can evade watermark detection for arbitrarily long text segments {\em within a reasonable edit distance budget, without using an unwatermarked LLM}. Our evaluation compares SCTS and existing representative attack methods over a series of edit distance budgets. We conclude that across various settings, our approach is superior in reducing AUROC (a proxy for detection success) to less than $0.5$ on two LLMs and two watermarking schemes. Additionally, and most importantly, our approach is theoretically grounded, with a comprehensive analysis. 

In summary, our main contributions are:
\begin{enumerate}
\itemsep0em
\item We theoretically show how existing color-agnostic methods can only ``dilute'' watermarks, and can not evade detection when the watermarked text is sufficiently long (\S~\ref{sec:building}). Empirically, we show that existing methods fail to evade watermarks within reasonable edit distance budgets (\S~\ref{section:results}).
\item We propose the first ``color-aware'' attack method by prompting the LLM for (a seemingly) random generation to obtain color information, and replace the green tokens/words with red tokens/words. We then analyze the working mechanism of our attack and estimate its efficacy and costs under reasonable edit distance constraints (\S~\ref{sec:our}).
\item We conduct extensive experiments on \texttt{vicuna-7b-v1.5-16k}~\citep{zheng2023judging} and \texttt{Llama-2-7b-chat-hf}~\citep{touvron2023llama} with different hashing strategies. We show that under the same edit distance budget, our attack is more effective in evading watermarks than previous methods (\S~\ref{section:results}).
\end{enumerate}

\section{Background and Related Work}
\label{sec:background}

\subsection{Background: Text Watermarking}

We utilize the work of~\citet{kirchenbauer2023watermark} as an exemplar approach; it involves perturbing the logits, so as to guide generation towards specific tokens. Their approach works as follows: an input sample $\mathbf{w} = w_{-s+1}, w_{-s+2}, \cdots, w_{0}$ of $s$ words is tokenized to obtain $s'$ tokens. These are then fed to the the large language model (LLM) which generates $T$ tokens as the response. Note that tokens are generated one by one. More specifically, before the generation of every output token, $c$ tokens (where $c$ is known as the context size) preceding that are used to seed a pseudo-random (hash) function which is used to divide the vocabulary space of size $|V|$\footnote{This is the vanilla left hash. A preferred variant is to also use the token being generated as an input to the hash. This is called self hash, and this is empirically more robust to watermark removal attacks.}, dynamically, into a green and red list, where the size of the green list is $\gamma |V|$. Based on re-weighting green tokens using an offset $\delta$ (more details in \S 3 of~\citet{kirchenbauer2023watermark}), the approach prioritizes the selection of a green list token as the next candidate. This process is repeated to generate all $T$ output tokens. The aim of their approach is to ensure that the actual number of green tokens in the generation ($||T||_G$) is higher than the expected number of green tokens ($\gamma T$). 
Post-hoc, this is verified using a statistical test:
$$z = \frac{\|T\|_G - \gamma T}{\sqrt{T \gamma (1 - \gamma)}}$$
The text is labeled as watermarked if and only if $z$ is greater than some threshold. 

\subsection{Attacks against Watermarks}

Most, if not all attacks involve replacing a subset of words. We describe two strategies below.

\noindent{\bf Strategy 1. Paraphrasing:} These attacks aim to replace a group of words with semantically similar counterparts. This can be done directly using a specialized LLM~\citep{krishna2023paraphrasing, sadasivan2023can}, word-level substitutions~\citep{shi2023red}, or translation (to another language and back)~\citep{christ2023undetectable}. For example, the recursive paraphrasing (\textsc{Rp}) approach~\citep{sadasivan2023can} 
paraphrases (up to 5 iterations) the watermarked text using a dedicated, unwatermarked paraphrasing model. 

\noindent{\bf Strategy 2. Prompting:} Another class of attacks involves carefully designing prompts to guide the model to generate text that evades detection~\citep{lu2023large}, or to guide the LLM to generate low-entropy text which is hard to watermark. As an example of the first category,~\citet{lu2023large} propose SICO-Para (\textsc{Sico}), where the LLM is tasked with generating features of human-written text. Using such features and human-written style examples, it can guide the LLM to augment the AI-generated text to be more human-like. \textsc{Sico} alternatively performs sentence and word-level updates, to greedily minimize the probability of detection using a proxy. As such a proxy is usually not a watermark detector and focuses on the semantics, it would not help when attacking watermarks.

In our work, we focus on \textsc{Rp} and \textsc{Sico} as representative baselines. 

\noindent\underline{\bf Limitations of Current Approaches:} Both approaches are {\em color-agnostic} i.e., meaning that the new fragments introduced by the attacker will be independent of the color of existing fragments (i.e., will have the same green token ratio as human-written text), while a large number of old fragments (which are mostly green) are also preserved~\citep{kirchenbauer2023reliability}. Consequently, the resulting text will still mostly comprise of green tokens, when we consider both existing and ``added'' fragments. This introduces natural tensions. Most importantly, both approaches are ineffective when there are constraints placed on the number of permitted edits. For example, we see that both \textsc{Sico} and \textsc{Rp}'s AUROC is still $\geq$ $0.86$ under the most relaxed $0.5$ normalized edit distance (i.e., editing $50\%$ of words of the watermarked text)\footnote{AUROC describes the distinction of the $z$-scores for (attacked) watermarked text and unwatermarked text and it is more comprehensive than Attack Success Rate (ASR).}.

\subsection{Threat Model}

Recall that the robustness of watermarks is defined as their tolerance to edits post-hoc. Prior work~\citep{lu2023large} evaluates robustness by making unrealistic assumptions about the (robustness) adversary's capabilities. They assume that the adversary has access to a version of an LLM without a watermarking algorithm. In our work, we do not make this strong assumption. Like other prior approaches~\citep{zhang2023watermarks, sadasivan2023can}, we assume:
\begin{enumerate}
\itemsep0em
\item API access to the watermarked model, using which we can issue input prompts, and observe the generated responses. 
\item The watermarked model is aligned, and capable of following instructions provided. 
\item Knowledge of the context size $c$.
\item No knowledge of other watermarking hyperparameters, like $\gamma$, temperature $t$, and $\delta$. 
\item Access to some model (not necessarily an LLM) capable of generating word substitution candidates. This model can be watermarked.
\end{enumerate}

We believe assumption 2 is realistic, given how most models in the status quo are instruction fine-tuned, and trained using reinforcement learning with human feedback~\citep{lambert2022illustrating}. We also stress that assumptions 2 and 3 are not strict.

\section{The Building Blocks}
\label{sec:building}

We will first show some properties of the watermarking efficacy as a function of output (generation) length, and use this to explain why existing color-agnostic attack methods fail for sufficiently long watermarked text. Then, we focus on substitution-based attacks and formalize the color-agnostic baseline. But before we begin, we outline the assumptions and definitions we make throughout this section.

\noindent{\bf Assumption 1:} We consider the left hash for ease of exposition. As stated earlier, the left hash is one where the green list for the current token is obtained by hashing $c$ tokens counting backward from the current token (not including it). 

\noindent{\bf Assumption 2:} Each (generated) token from watermarked LLM is green with constant probability $p$ in watermarked text generation, i.i.d, s.t. $ \gamma < p < 1$. In the generated output, the number of $c$-grams is $T-c$. 

\begin{definition}[{$c+1$-gram}]  
This is the set of tokens used for color testing. It includes the token whose color is to be checked and the $c$ tokens before it. The color of a $c+1$-gram refers to the color results when this $c+1$-gram is sent to the detector. 
\end{definition}

\begin{definition}[Effective length $T_e$] This is the number of $c+1$-grams used for color testing i.e., $T_e=T-c$
\end{definition}

\begin{definition}[Detection threshold $z_{th}$] This is the threshold used in detection. The detector outputs ``watermarked'' $\iff z > z_{th}$. 
\end{definition}

\begin{definition}[Critical length $T_c$] This is the value of $T_e$ such that $\mathbb{E}[z]=z_{th}$. This can be thought of as the effective length needed for successful detection, and may or may not exist.
\end{definition}

\begin{definition}[Average green probability $q$]: This is the average probability for a $c+1$-gram being green for any given arbitrary sample, which can be watermarked or non-watermarked, attacked or non-attacked. By definition, $\mathbb{E} [\|T\|_G]=qT_e$.
\end{definition}

\subsection{Theorem: Watermark Strength}

\begin{theorem}
\label{theorem}
$\mathbb{E}[z]$ is proportional to $\sqrt{T_e}$. To elaborate, first assume the colors of $c+1$-grams are independent. Then, we have:

\begin{itemize}
\itemsep0em
\item For \(q < \gamma\), the probability of detection as ``watermarked'' converges exponentially to 0 with respect to \(T_e\).
\item For \(q > \gamma\), the probability of being detected as ``unwatermarked'' converges exponentially to 0 with respect to \(T_e\).
\end{itemize}

Furthermore, if the color for different $c+1$-grams is green, is i.i.d., then:

\begin{itemize}
\itemsep0em
\item For \(q \neq \gamma\), tighter bounds are applicable in comparison to the scenarios described above in the independent case.
\item For \(q = \gamma\), the probability of being detected as ``unwatermarked'' converges to a constant determined by $z_{th}$.
\end{itemize}
\label{ref:thm1}
\end{theorem}

The proof of the above is in Appendix~\ref{app0}.

\subsection{Why Existing Methods Fail For Long Text}

As a warm-up, let us consider text that is not attacked. Applying Theorem~\ref{ref:thm1}, where $q=p > \gamma$ (for the i.i.d. case), we can see that the false positive rate converges to $0$ exponentially w.r.t. $T_e$.

\noindent{\bf {Candidate Attack 1}.} There are attacks that design prompts to guide the watermarked LLM to generate low-entropy text to evade detection. It is hard to incorporate watermarks in such (generated) text segments. But even if one can circumvent this challenge, we will see that it may only dilute the watermark. To this end, assume that the attacker can reduce $p$ to $p'$, such that $p > p'>\gamma$, i.e., the $c+1$-grams are i.i.d green with probability $p'$. Note, however, that when $p'>\gamma$, we can apply theorem \ref{ref:thm1} where we set $q=p'>\gamma$ (for the i.i.d. case as before); the false positive rate converges to $0$ exponentially w.r.t. $T_e$.

\begin{tcolorbox}
\noindent{\bf Conclusion.}
This attack would fail for sufficiently long watermarked text, and it can only dilute the watermark.
\end{tcolorbox}

\noindent{\bf {Candidate Attack 2.}} Here, the attacker uses post-processing methods to evade detection. Existing paraphrasing-based methods~\citep{lu2023large, krishna2023paraphrasing} fall into this category. Since the post-processing is color-agnostic, we assume the newly generated $c+1$-grams are green with i.i.d. probability $\gamma$, and the number of such {\em new $c+1$-grams} is $T_{new}$. Also note that post-processing attacks such as paraphrasing are statistically likely to leak $n$-grams or even longer fragments of the original text~\citep{kirchenbauer2023reliability}. We call the leaked segments {\em ``old'' $c+1$-grams}. These old $c+1$-grams are green with i.i.d. probability $p$, and the number of such old $c+1$-grams is $T_{old}$. The existence of two classes of $c+1$-grams suggests that when detecting the watermark from the attacked text, $T_e=T_{old}+T_{new}$. The ratio of leaked $c+1$-grams is $r_o=\frac{T_{old}}{T_{old}+T_{new}}$, and is lower bounded by constant $r>0$. Finally, the colors for different $c+1$-grams categories are independent. 

\begin{definition} We define
\label{ToTn}
\begin{itemize}
    \item $\|T\|_G^{old}$, the random variable for the number of ``old'' $c+1$-grams which are green.
    \item $\|T\|_G^{new}$, the random variable for the number of ``new'' $c+1$-grams which are green.
\end{itemize}
So $\|T\|_G^{old}\sim B(T_{old},p)$, $\|T\|_G^{new}\sim B(T_{new},\gamma)$, $\|T\|_G=\|T\|_G^{old}+\|T\|_G^{new} $.
\end{definition}

Applying the independent case of the Theorem~\ref{ref:thm1} ($q > \gamma$), we can see that the probability of evading detection exponentially converges to $0$ w.r.t. $T_e$.

\begin{tcolorbox}
\noindent{\bf Remark 1.} In paraphrasing attacks, we can safely assume that attack makes the text longer. 

\noindent{\bf Remark 2.} Even if attack 1 and attack 2 are combined, i.e., the attacker can weaken the watermark and do color-agnostic post-processing, then it would still fail for sufficiently long text. To see this, just let $p=p'>\gamma$ in the above derivation.

\noindent{\bf Conclusion}
This attack would fail for long enough generated text when a non-zero ratio of $c+1$-grams is preserved. 
\end{tcolorbox}

\section{Our Approach: Self Color Testing} 
\label{sec:our}

We first introduce some notation. The watermarked sentence (i.e., the output of the LLM) is denoted $\mathbf{w}=\{w_1,w_2,\cdots, w_T\}$. In this sentence, the word being substituted is denoted {$w_b$}. The candidate substitution is $w'_b$. Finally, {$\mathbf{w}_b^c$} denotes the context of $w_b$, which determines the green list at location $b$ i.e., $\mathbf{w}_b^c$ is the $c$ words before $w_b$ (assuming left hash)

\vspace{1mm}
\noindent{\bf Step 1. Self Color Testing:} The secret to a successful attack against watermarked LLMs is to have color information. However, obtaining this is not straightforward. We leverage the insight that aligned and instruction fine-tuned models are compliant with user instructions. Thus, we can prompt the LLM (at temperature $t=0$) to generate seemingly random strings (in a deterministic manner) with customized input prefixes, and infer color information from the frequency of outputs generated (abstracted by the ``\texttt{Count}'' method in Algorithm~\ref{ref:alg1}). We focus our discussion on the word level to avoid encoding issues and incomplete word biases. Algorithm~\ref{ref:alg1} contains all relevant details.

\begin{algorithm}
\footnotesize
\SetAlgoLined 
\textbf{Input:} $w_b$, $w_b'$, $\mathbf{w}_b^c=\{w_{b-c}\ldots w_{b-1}\}$, $p_{th}$\;
\textbf{Output:} Test result in $\{\texttt{GR,RG,S}\}$. \texttt{GR} (\texttt{RG}) means $w_b$ is green (red) while $w_b'$ is red (green), \texttt{S} means $w_b$ and $w_b'$ are of the same color\;

$p=\text{MakePrompt}(w_b, w_b', \mathbf{w}_b^c)$\;
$o=\text{LLM}(p)$\;
$c_b, c_b' = \texttt{Count}(o, \mathbf{w}_b^cw_b), \texttt{Count}(o, \mathbf{w}_b^cw_b')$\;
$p_t=\chi^2\text{-test}(c_b, c_b')$\;
  \If{$p_t \geq p_{th}$}{
    \Return \texttt{S}\;
    }
    \uElseIf{$c_b>c_b'$}{
    \Return \texttt{GR}\;
    }
    \Else{
    \Return \texttt{RG}\;
    }
\caption{Self Color Testing (SCT)}
\label{ref:alg1}
\end{algorithm}

To explain the intuition behind the algorithm, consider the following example prompt for color testing. Here, note that $(w_b$=\texttt{includes}, $w'_b$=\texttt{contains}, and $\mathbf{w}_b^c$=\texttt{kernel})

\begin{tcolorbox}
\footnotesize
\texttt{Choose two phrases ( kernel includes, kernel contains ), and generate a long uniformly random string of those phrases separated by ";". Previous phrases should have no influence on future phrases: kernel includes; kernel contains; kernel includes; kernel contains; kernel contains; kernel includes; kernel contains; kernel contains; kernel contains; kernel contains; kernel contains; kernel contains; kernel contains; kernel contains; kernel includes; kernel contains; kernel contains; kernel includes; kernel contains; kernel contains;} 
\end{tcolorbox}
When the above prompt is processed by the LLM, we check the frequency of ``\texttt{kernel includes}'' and ``\texttt{kernel contains}'' in the response. Assume the token before ``\texttt{includes}'' and ``\texttt{contains}'' are the same in the encoded ``\texttt{kernel includes}'' and ``\texttt{kernel contains}''. Since the model is deterministic (temperature $t=0$), and follows the instructions in the prompt, the model will always generate the green token if $w_b$ and $w_b'$ are of different colors. Thus, SCT has a perfect recall on \texttt{GR} and \texttt{RG}. For the \texttt{S} scenario, there can be border cases, but in realistic applications ($t>0$), two tokens of the same color will have the same probability of being generated.

\vspace{1mm}
\noindent{\bf Step 2. SCT Substitution:} We use color-testing to test different candidates to ensure that the green token is substituted by a red one. Note that (a) \texttt{Generate\_candidates}$(\mathbf{w}',i)$  generates $k$ substitution candidates (different from the original) for $\mathbf{w}'$ at index $i$, and (b) \texttt{Substitute}$(\mathbf{w}',i, w_i')$ updates the $i$-th word of $\mathbf{w}'$ with $w_i'$.
The choice of step size $2$ upon successful substitution is heuristic: consider the left hash $c=1$ case. Suppose we just substituted $w_i$ with $w_i'$, this will also change the green list at $w_{i+1}$. It is reasonable to assume the probability of being green for $w_{i+1}$ is reduced to $\gamma$ (from $p$) as we did not check its color before substitution. Consequently, continuing to substitute $w_{i+1}$ is less likely to be successful, as $w_{i+1}$ has a lower probability of being green, which in turn will lead to more computation.

\begin{algorithm}
\footnotesize
\SetAlgoLined 
\textbf{Input:} $\mathbf{w}$, \\
\textbf{Output:} $\mathbf{w}'$\\
\textbf{Initialization:} $i=c$, $\mathbf{w}'=\mathbf{w}$\\
\While{$i \leq T$}{
    $w_{i,1}',\ldots,w_{i,k}' = \texttt{Generate\_candidates}(\mathbf{w}', i)$\\
    $success=\text{False}$\\
    \For{$j=1$ \KwTo $k$}{
        \If{$\text{SCT}(w_i, w_{i,j}', \{w_{i-c}\ldots w_{i-1}\}) == \texttt{GR}$}{
            $\mathbf{w}'=\texttt{Substitute}(\mathbf{w}',i,w_{i,j}')$\\
            $success=\text{True}$\\
            break
        }
    }
    \eIf{success}{
        $i\gets i+2$\\ (advance by 2 for economic substitution)
    }{
        $i\gets i+1$\\
    }
}
\caption{SCTS Algorithm}
\end{algorithm}

\vspace{1mm}
\noindent{\bf Budget Enforcement:} To enforce the budget, we add checks after each substitution. More details are in Appendix~\ref{Budgets}.

\subsection{Analysis}

To simplify our analysis, we assume that each word corresponds to a single token, and that the candidate generation process is color-agnostic i.e., every candidate is green with i.i.d probability $\gamma$. We focus on $c=1$ left hash for simplicity.

\noindent{\bf Expected Green Ratio.}
Let $q(T_e)$ be the average green probability. When the number of candidates in each substitution attempt $k>1+\log_\gamma \frac{1-p}{p(1-\gamma)}$, we have
\begin{align}
q(T_e)\leq \max \{\frac{\gamma}{2}, \frac{\gamma^k p}{1-p+\gamma^k p}\} < \gamma
\label{eq:qTeBound}
\end{align}
\begin{equation}
\lim_{T_e \to \infty} q(T_e) = \gamma - \frac{\gamma(1-p\gamma^{k-1})}{1+(1-\gamma^k)p} < \gamma
\label{eq:limit_expression}
\end{equation}

Using similar techniques as theorem~\ref{ref:thm1}, we have 
\begin{align}
\log_e\Pr(z > z_{th}) < -(\gamma-q(T_e))^2\Bigg(\sqrt{T_e} \nonumber +\frac{\sqrt{\gamma(1-\gamma)}z_{th}}{\gamma-q(T_e)}\Bigg)^2
\label{eq:ProbabilityBound}
\end{align}
for $q(T_e)<\gamma$, which holds for large enough $k$ and/or $T_e$. Thus, the probability of failing to evade the watermark converges to $0$ exponentially.
This means that, in expectation, our method can reduce the average green probability to less than $\gamma$, and is amenable to arbitrarily long text. More details are presented in Appendix~\ref{SCTS}.

\vspace{1mm}
\noindent{\bf LLM Calls.} We also analyze the number of LLM calls needed for our approach in Appendix~\ref{app1}. Our method is reasonably fast, with $\mathcal{O}(1)$ w.r.t. $k$, and $\mathcal{O}(T_e)$ w.r.t. $T_e$ in expectation.


\begin{figure}[htbp]
\centering
\includegraphics[width=\linewidth]{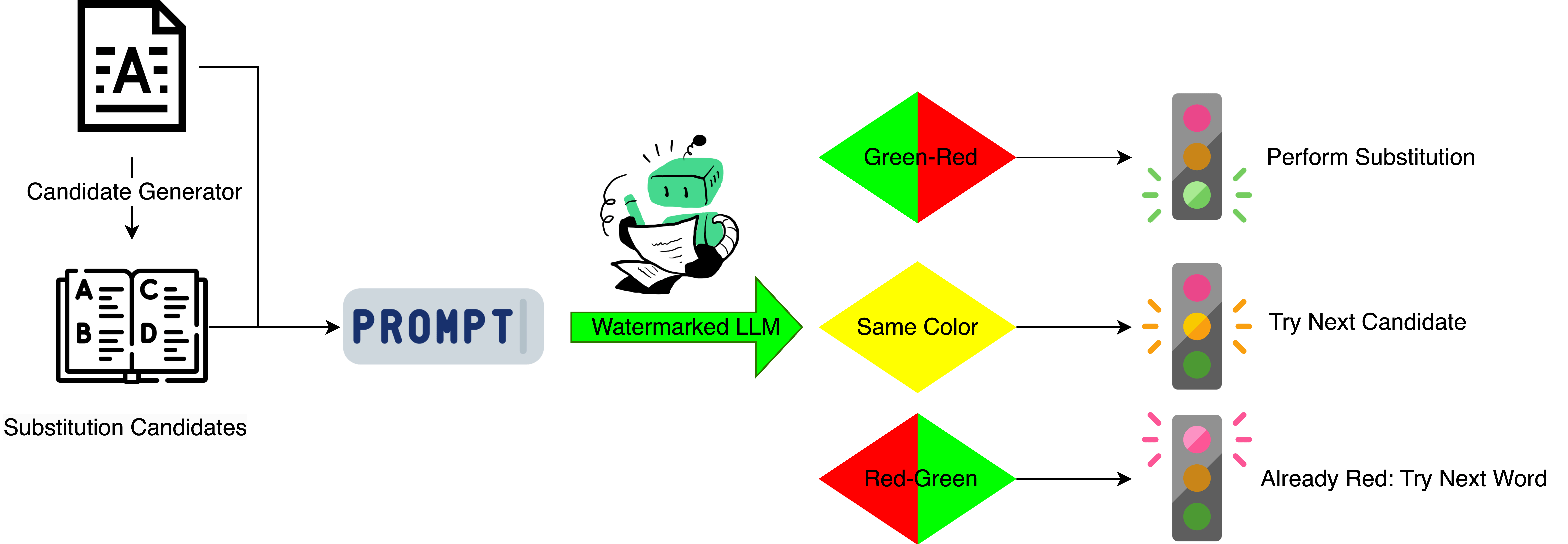}
\caption{Illustration of one substitution in SCTS for simplicity. Take different actions depending on the frequency in the SCT test.\vspace{-3mm}}
\label{fig:master}
\end{figure}

\section{Experimental Setup}

\noindent{\bf Constraints:} We consider word level edit distance normalized by the number of words for the watermarked text. The constraints we use are from 0.05 to 0.5, with intervals of 0.05. 

\subsection{Baseline Methods}

\noindent{\bf 1. \textsc{Rp}.} To enforce edit-distance constraints, we check if the paraphrased sample is within the constraint. If not, we use the unparaphrased sample as the result. The other settings are as suggested by the official implementation of~\citet{sadasivan2023can}. \textsc{Rp}$i$ denotes $i$ cycles of recursive paraphrasing; $i=\{1,2,3,4,5\}$.

\noindent{\bf 2. \textsc{Sico}.} We focus on the SICO-para variant~\citep{lu2023large}. To enforce edit constraints, we add a description of the constraint in the prompt and only include compliant samples. The detailed prompt is in Appendix~\ref{app:additional_exp}. 
The other settings are kept the same as in~\citet{lu2023large}. The training samples are also the same as that used in~\citet{lu2023large}.

\noindent{\bf 3. \textsc{Rb}.} This is a baseline for substitution without color testing. It uses the same candidate generation but without SCT. This is used to show if the color information from SCT is useful.

\subsection{Models}

We use \texttt{vicuna-7b-v1.5-16k} and \texttt{Llama-2-7b-chat-hf} as watermarked models. For the candidate generation model, we use the HuggingFace pipeline using \texttt{distilroberta-base}~\citep{Sanh2019DistilBERTAD}. We have $k=5$ substitution candidates for each word.

\subsection{Watermarking Scheme}

We consider two schemes.

\noindent{1.} \underline{UMD} by~\citet{kirchenbauer2023watermark}: Here, we consider both left hash (or Min-LeftHash) and self hash (or Min-SelfHash) (more details in~\citep{kirchenbauer2023reliability}), and context size of $c=1,2,4,8$. As suggested by the authors, we set $z_{th}=4$, $\gamma=0.25$, and $\delta=2$. For efficiency, we batched the self hash. We consider two models: \texttt{vicuna-7b-v1.5-16k} and \texttt{Llama-2-7b-chat-hf} . 

\noindent{2.}  \underline{Unigram} by~\citet{zhao2023provable}: This is a variant of UMD's SelfHash with $c=1$, but {\em whose green list is fixed}. Here, we set $z_{th}=4$, $\gamma=0.5$, and $\delta=2$ (based on their released code\footnote{We reduce $z_{th}$ from $6$ to $4$ to make Attack Success Rate (ASR) more representative}). We consider only one model: \texttt{vicuna-7b-v1.5-16k} \footnote{For UMD self hash $c=1$ and Unigram, we slightly modify the SCTS prompt (Appendix~\ref{prompt}), and it will always advance by 1 regardless if the substitution succeeds.}.

\noindent{\bf Other Parameters}
The $p$-value threshold for the $\chi^2$ test in SCT is set to $0.01$. \\

More details about the datasets and metrics we calculate are presented in Appendix~\ref{app:additional_exp}.
\section{Experiment Results}
\label{section:results}

Through our evaluation, we wish to answer the following questions: 
\begin{enumerate}
\itemsep0em
\item Is SCTS more effective than previous methods? \item What budget is required for a successful attack?
\item Does the above hold for different target watermarked models?
\end{enumerate}
    
We observe that: 
\begin{enumerate}
\itemsep0em
\item SCTS is consistently more effective for different values of $c$, hashing schemes, and watermarking schemes, while preserving output semantics (\S~\ref{effective}).
\item Normalized edit distance of $0.25-0.35$ is sufficient for SCTS, while other attacks need more (\S~\ref{budget}).
\item Attack Success Rate (ASR) is heightened for models which are aligned and instruction fine-tuned (\S~\ref{diffmodel}).
\end{enumerate}

\begin{figure}[htbp]
\centering
\includegraphics[width=\linewidth]{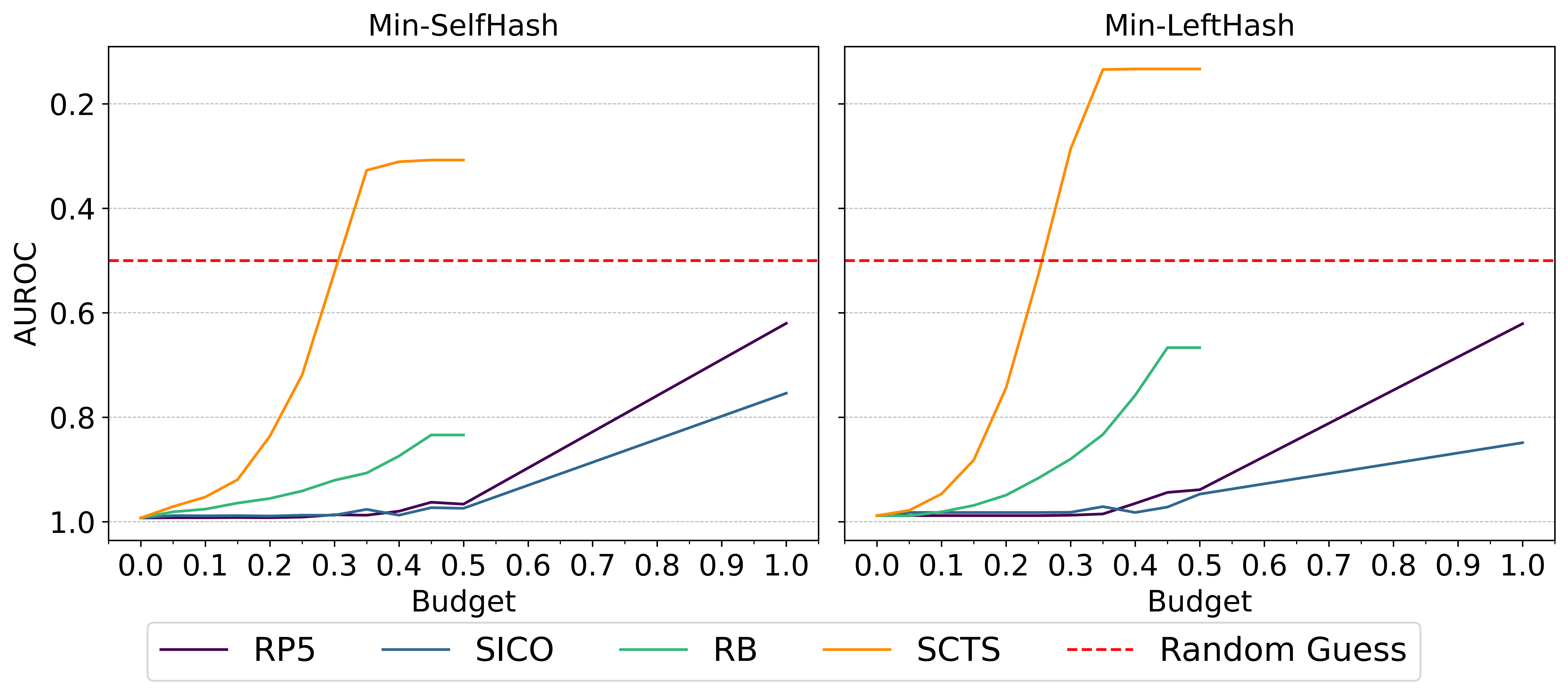}
\caption{AUROC for \texttt{vicuna-7b-v1.5-16k}, 50 samples, UMD watermarking, $c=4$. The orange curve (SCTS) is consistently and significantly above other baselines, and it is the only one cross $0.5$.\vspace{-3mm}}
\label{fig:AUROC}
\end{figure}

For most metrics, we visualize the $c=4$ case of UMD due to space constraints. Detailed results (for both UMD and Unigram) are presented in Appendix~\ref{tables}, and highlight the same trends. In particular, we observe that Unigram is more robust than UMD, but still susceptible to SCTS. For \textsc{Rp}, we only visualize \textsc{Rp}5 as it is the strongest attack. 

\subsection{Is SCTS Effective?}
\label{effective}

\begin{figure}[htbp]
\centering
\includegraphics[width=\linewidth]{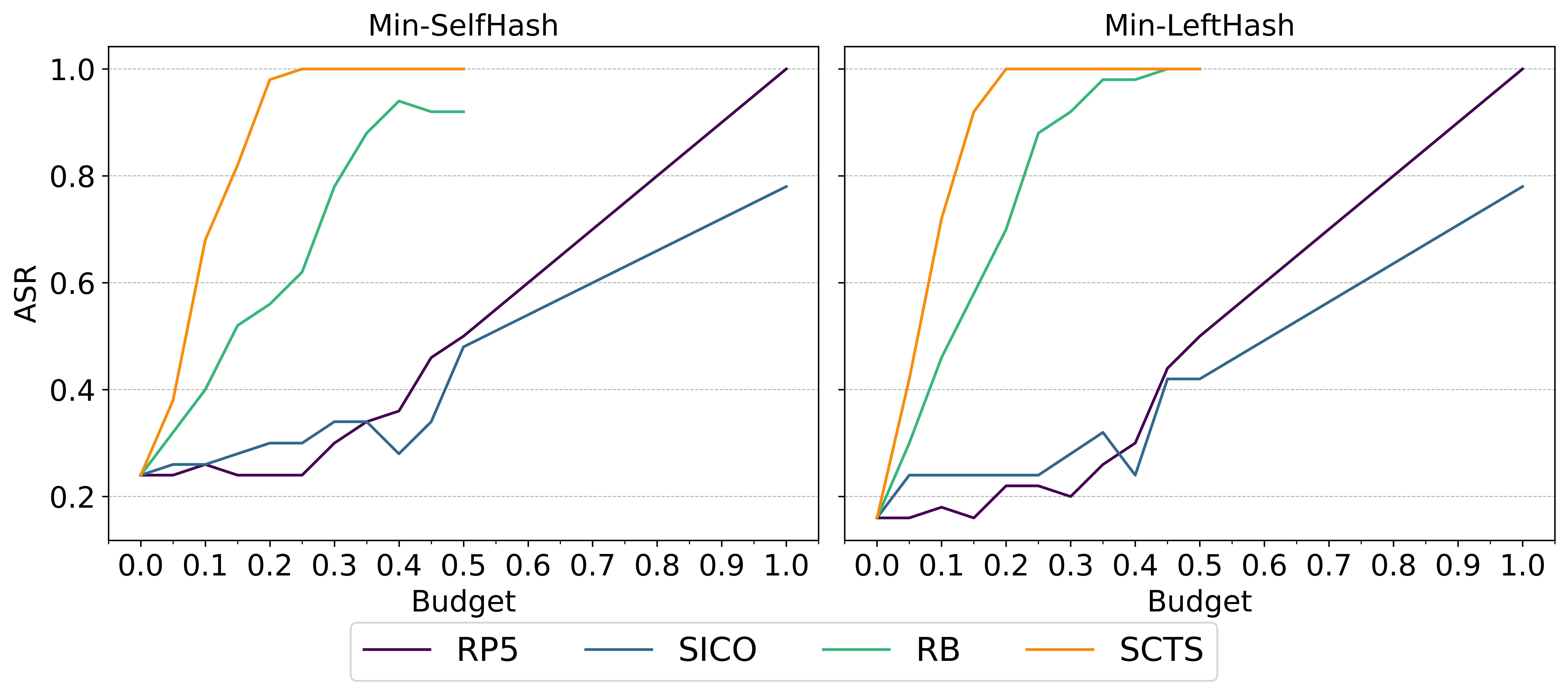}
\caption{ASR for \texttt{vicuna-7b-v1.5-16k}, 50 samples, UMD, $c=4$, $z_{th}=4$. SCTS (orange) can significantly evade more detection under the same budget than other baselines.\vspace{-3mm}}
\label{fig:SR}
\end{figure}

\underline{\em Yes, it is.} Figure~\ref{fig:AUROC} and~\ref{fig:SR} shows the performance of the attack on AUROC and detection success respectively over different edit budgets. We can see that SCTS is consistently more effective.

At relatively high budgets like $0.35$, SCTS can reduce the AUROC to less than $0.5$, which means that the $z$-score (used for detection) is, on average, more negative. This in turn corresponds to the scenario where $q < \gamma$ in equation~\ref{eq:limit_expression}, despite some of the assumptions we made not holding. 

\textsc{Sico} and \textsc{Rp}, in contrast, fail to evade detection at most budgets, and are even worse than \textsc{Rb}. Even though they work reasonably well when they are unconstrained, there are still a few detectable samples. Also, their AUROCs are still in the range of 0.6 to 0.9, suggesting that the watermark is generally only diluted and can be detected for longer text. 

\noindent{\bf Impact of $c$ and hashing scheme.} Both figures show that self hash is generally more robust than the left hash, especially for \textsc{Rb} and SCTS. This is consistent with the findings in~\citet{kirchenbauer2023reliability}, and also holds for $c=2,4,8$. Also, smaller $c$ is generally more robust from our experiment results in the Appendix~\ref{tables}, Table~\ref{tab:A1},~\ref{tab:D1},~\ref{tab:A3},~\ref{tab:D3}, consistent with the findings of~\citet{kirchenbauer2023reliability}. Nevertheless, smaller $c$ comes with a higher risk of leaking the green $c+1$-grams to an attacker, more loss in generation quality~\citep{kirchenbauer2023reliability}, lower $z$ and successful detection rates as shown in our experiments.

\begin{table*}[!t]
\centering
\caption{Semantic similarity for \texttt{vicuna-7b-v1.5-16k}, 50 samples, UMD watermarking, $c=4$. SCTS successfully preserves semantics.}
\label{tab:similarity}
\begin{adjustbox}{width=\textwidth}
{\small
\begin{tabular}{@{}lcccccccccccc@{}}
\toprule
\textbf{Hashing} & \textbf{Method} & \textbf{0.05} & \textbf{0.1} & \textbf{0.15} & \textbf{0.2} & \textbf{0.25} & \textbf{0.3} & \textbf{0.35} & \textbf{0.4} & \textbf{0.45} & \textbf{0.5} & \textbf{1 (Unconstrained)} \\
\midrule
\multirow{4}{*}{Left} 
 & \textsc{Rp}5 & 1.0000 & 0.9979 & 0.9978 & 0.9986 & 0.9922 & 0.9894 & 0.9881 & 0.9711 & 0.9586 & 0.9467 & 0.5436 \\ 
 & \textsc{Sico} & 0.9984 & 0.9988 & 0.9906 & 0.9889 & 0.9884 & 0.9886 & 0.9934 & 0.9910 & 0.9845 & 0.9600 & 0.7104 \\ 
 & \textsc{Rb} & 0.9862 & 0.9728 & 0.9538 & 0.9394 & 0.9203 & 0.9008 & 0.8854 & 0.8732 & 0.8546 & 0.8546 & - \\ 
 & SCTS & 0.9878 & 0.9743 & 0.9579 & 0.9366 & 0.9198 & 0.9018 & 0.8842 & 0.8832 & 0.8832 & 0.8832 & - \\ 
\midrule
\multirow{4}{*}{Self} 
 & \textsc{Rp}5 & 0.9998 & 0.9987 & 0.9988 & 0.9916 & 0.9931 & 0.9841 & 0.9709 & 0.9582 & 0.9488 & 0.9416 & 0.5464 \\ 
 & \textsc{Sico} & 0.9979 & 0.9999 & 0.9925 & 0.9880 & 0.9614 & 0.9869 & 0.9811 & 0.9857 & 0.9722 & 0.9674 & 0.6439 \\ 
 & \textsc{Rb} & 0.9847 & 0.9686 & 0.9507 & 0.9330 & 0.9141 & 0.8964 & 0.8801 & 0.8611 & 0.8459 & 0.8459 & - \\ 
 & SCTS & 0.9850 & 0.9673 & 0.9499 & 0.9285 & 0.9105 & 0.8891 & 0.8754 & 0.8737 & 0.8732 & 0.8732 & - \\ 
\bottomrule
\end{tabular}
}
\end{adjustbox}
\end{table*}

\noindent{\bf Semantic similarity.}
Another key factor for a successful attack is if it can preserve semantics. From Table~\ref{tab:similarity}, notice our method successfully preserves semantics during the attack, with a mean cosine similarity $0.8832$ and $0.8732$ at $0.4$ budget (for left and self hash respectively), which is comparable to \textsc{Rp}1. Results for the Unigram watermark are slightly lower, and more details are presented in Table~\ref{tab:S2} in Appendix~\ref{full}. We believe that semantics will be better preserved if we use more powerful substitute generators, or make larger substitutions (phrases vs. words as we currently do).


\begin{figure}[htbp]
\centering
\includegraphics[width=\linewidth]{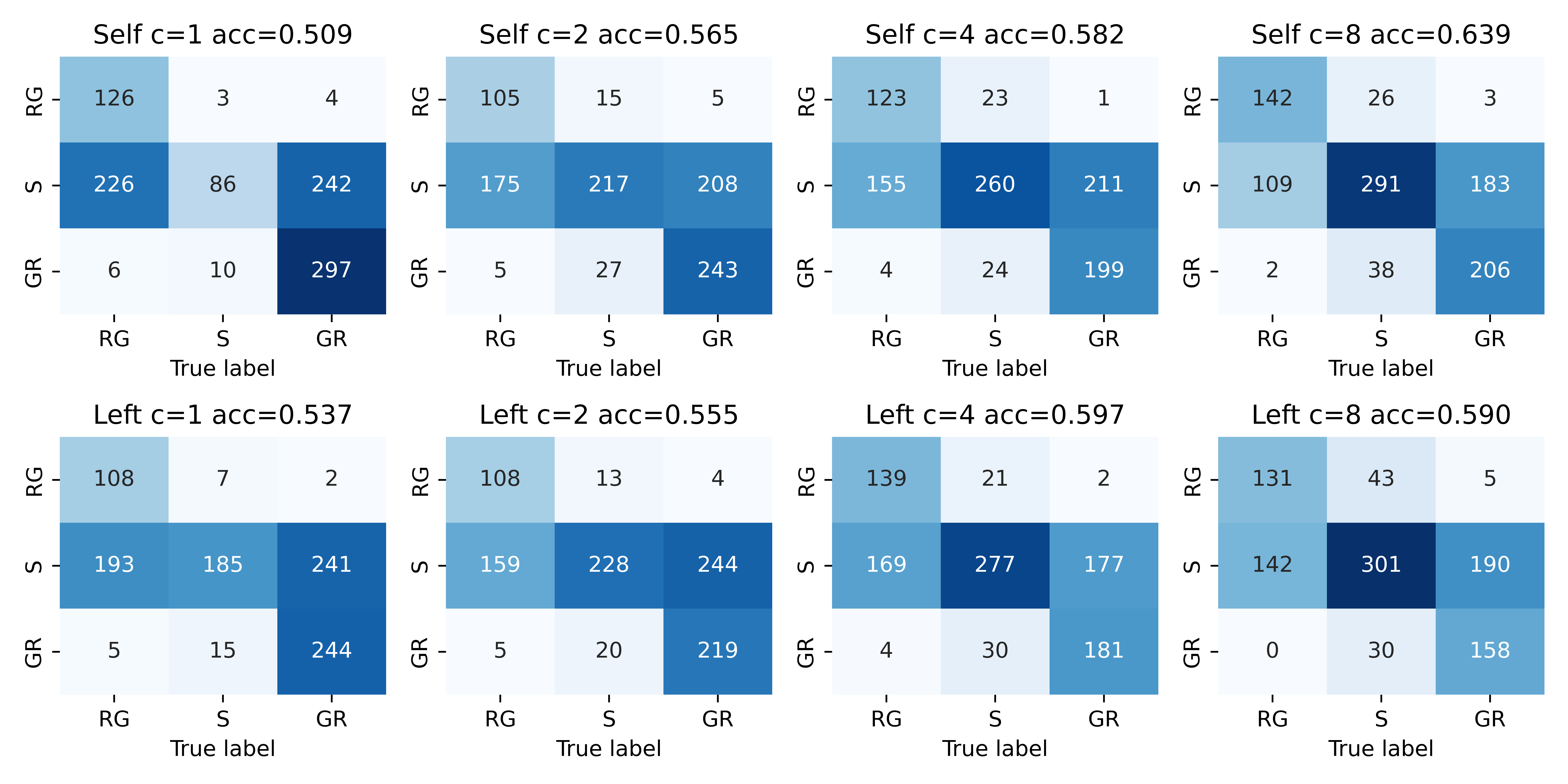}
\caption{Confusion matrix and accuracy for SCT over 1000 samples for \texttt{vicuna-7b-v1.5-16k}, UMD. Accuracies are at least $0.5$ for all $c$ and hashing.  \vspace{-3mm}}
\label{fig:SCT}
\end{figure}

\noindent{\bf SCT Accuracy.} The key factor for our success is the accuracy of SCT. From Figure~\ref{fig:SCT}, we see that the accuracy for SCT is $\geq 0.5$. This is significantly higher than $0.33$, a rough random baseline estimate for a three-class classification problem, suggesting the effectiveness of SCT. This is particularly impressive given that~\citet{kirchenbauer2023watermark}'s coloring scheme is designed to be hard to identify.

\subsection{Attack Budget}
\label{budget}

For samples we evaluated, SCTS is successful at evading detection with as low as $0.25$ normalized edit distance. Albeit \textsc{Rb} also maximizes ASR at higher budgets, only SCTS is able to consistently reduce AUROC to less than 0.5. The trend is consistent for all values of $c$ considered. 

However, note that SCTS saturates in the $0.35-0.45$ interval. This is because the current implementation only scans the sentence once when performing substitutions, and the ASR for each attempt can be lower than the theoretical result. Determining techniques to scan the sentence multiple times to increase residual replacement success is subject to future research.

\subsection{Different Watermarked Models}
\label{diffmodel}


Notice that the success of SCT is implicitly connected to the ability of the model to follow instructions. We observed that \texttt{Llama-2-7b} (model that is not instruction fine-tuned) would frequently fail to follow the prompt. For example, when the prompt is to let it generate phrases from \texttt{kernel includes} and \texttt{kernel contains}, it would probably generate several of these phrases and then begin to generate \texttt{includes} and \texttt{contains} (in words only instead of phrases), whose green lists are different. Consequently, there are insufficient number of samples for the $\chi^2$ test leading to bias in SCT towards \texttt{S}. In this case, the success rate for each substitution attempt would drop, and SCTS would saturate at a lower budget, become slower, and less effective. On the other hand, the chat variant \texttt{Llama-2-7b-chat-hf} suffers less from this issue, while \texttt{vicuna-7b-v1.5-16k} (further fine-tuned from \texttt{Llama-2-7b-chat-hf}) is even better. We would argue that as models become more aligned and instruction fine-tuned, they will follow the prompts better in general, and SCTS will be more effective.


\section{Discussion and Open Questions}
\subsection{Limitations} 
One limitation of SCTS is that its efficiency can be improved, as shown in \ref{fig:R}, \ref{fig:L}. The color information is also limited to one pair due to an attacker can only prompt the black-box watermarked model. We assume the attacker knows $c$, while a workaround is possible for future work. SCT accuracy is also not very high so the color information is not that accurate. Lastly, SCTS is currently limited to UMD and its variants like Unigram.

\subsection{Open Questions}

\noindent{\bf Can SCTS be faster?}
One way is to store the color information already found to reduce repetitive color testing, with the risk of the accumulation of incorrect results and the cost of space. Such caching is more practical for small $c$.

\noindent{\bf Can one LLM query get more color information?}
 Currently, our color testing can only test one pair in each LLM query, and it can not distinguish if the two words/tokens are both red or green. 
More candidates for random generation can help with the cost of more undesired factors getting involved, like the increased complexity of the prompt, the reduced frequency count for each candidate, and the more complex cases in hypothesis testing. 

\noindent{\bf Unknown $c$}
Besides prompting the model to guess $c$ first, one way to use SCTS in this case is to use a large estimated $c$. We leave this for future work.

\noindent{\bf Can the accuracy of SCT be higher?}
 \citep{tang2023baselines} shows the complicated behavior when the LLM model is prompted to generate uniform random strings, which are far from uniformity and vary model by model. Such behavior makes our color testing sometimes inaccurate. One possible way is to have multiple variants, like exchanging the position of the new candidate and the old to do a second prompt, to improve accuracy with the cost of more computation.

\subsection{Harms}

Through this work, we propose an approach to circumvent text watermarking strategies. This has implications for spreading misinformation and purporting enhanced (nefarious) dual-use of LLMs. We hope that our findings can help design more robust watermarking techniques. 
\section{Conclusions}
Our study presents SCTS, an algorithm to evade watermark detection without using external LLMs. We demonstrate that SCTS can effectively eliminate watermarks from long texts using a straightforward algorithm. This approach reveals that specific prompting techniques can uncover and exploit private watermarking information, enabling evasion. We aim to inspire further research on developing more robust and secure watermarking schemes.

\bibliography{custom}
\bibliographystyle{acl_natbib}

\appendix

\onecolumn 
\newpage
\section*{Appendix}

\section{Proof of Theorem 1}
\label{app0}

\noindent{\bf Theorem 1:} $\mathbb{E}[z]$ is proportional to $\sqrt{T_e}$. \\

To elaborate, first assume the colors of $c+1$-grams are independent. Then, we have:

\begin{itemize}
\itemsep0em
\item For \(q < \gamma\), the probability of detection as ``watermarked'' converges exponentially to 0 with respect to \(T_e\).
\item For \(q > \gamma\), the probability of being detected as ``unwatermarked'' converges exponentially to 0 with respect to \(T_e\).
\end{itemize}

Furthermore, if the color for different $c+1$-grams is green, is i.i.d., then:

\begin{itemize}
\itemsep0em
\item For \(q \neq \gamma\), tighter bounds are applicable in comparison to the scenarios described above in the independent case.
\item For \(q = \gamma\), the probability of being detected as ``unwatermarked'' converges to a constant determined by $z_{th}$.
\end{itemize}

\noindent{\bf Proof:} Recall that $\|T\|_G$ is the number of green $c+1$-grams out of $T_e$ $c+1$-grams in the text, and 
$$z = \frac{\|T\|_G - \gamma T_e}{\sqrt{T_e \gamma (1 - \gamma)}}$$

For the expected $z$, recall $\mathbb{E} [\|T\|_G]=qT_e$ so that
$$\mathbb{E}[z] =\frac{q-\gamma}{\sqrt{\gamma(1-\gamma)}}\sqrt{T_e}$$
The proportionality ($\propto$) relationship is obvious. For $T_c$, simply let $\mathbb{E}[z]=z_{th}$ and solve for $T_e$ (results below).

\subsection{\bf Case 1:  i.i.d. case}
Let $D(a||q)$ be KL-divergence with base $e$ be defined as follows:
$$D(a||q)\coloneqq a \ln \frac{a}{q}+(1-a)\ln{\frac{1-a}{1-q}}$$

\begin{enumerate}
\item[S1.] If $q > \gamma $, then $\mathbb{E}[z]  \propto \sqrt {T_e}$ and 
$T_c=\frac{\gamma(1-\gamma)z_{th}^2}{(q-\gamma)^2}$

\noindent{\bf How?} Under assumption 2 that watermarked tokens are green with $i.i.d$. probability $p$, we have $\|T\|_G \sim B(T_e,q)$. From the Chernoff bound \citep{arratia1989tutorial}, for arbitrary $y\; s.t.\; 0 \leq y\leq T_e q$:
\begin{equation}
\Pr (\|T\|_G \leq y) \leq \exp (-T_e D(\frac{y}{T_e}||q)) \label{eq:Cbound}
\end{equation}
When $T_e \geq T_c$, take $y=z_{th}\sqrt{T_e\gamma(1-\gamma)}+\gamma T_e \leq qT_e$, we have:
\begin{equation*}
\begin{split}
  \Pr(z \leq z_{th}) 
  \leq \exp\left(-T_eD\left(\frac{\sqrt{\gamma(1-\gamma)} z_{th}}{\sqrt{T_e}}+\gamma || q \right)\right)
\end{split}
\label{eq:iid_bound_1}
\end{equation*}

Note that
$$
\lim_{T_e \to \infty} \frac{-T_eD(\frac{\sqrt{\gamma(1-\gamma)} z_{th}}{T_e}+\gamma||q)}{T_e}=-D(\gamma||q)
$$
 
$\implies$ Probability of being labeled ``not watermarked'' converges to 0 exponentially w.r.t. $T_e$.

\item[S2.] If $q=\gamma$, then $\mathbb{E}[z] = 0$, and $T_c$ does not exist. 

\noindent{\bf How?} According to the De Moivre–Laplace theorem \citep{dunbar2011moivre}, as \(T_e \to \infty\), the distribution of \(z\) approaches the standard normal distribution \(N(0,1)\). This convergence allows us to use the properties of the standard normal distribution to estimate probabilities related to \(z\). Specifically, the probability of \(z\) being detected as ``watermarked'' when exceeding a threshold \(z_{th}\) can be expressed as:
$$
\lim_{T_e\to \infty} \Pr(z>z_{th}) = 1-\Phi(z_{th})
$$
where \(\Phi(z_{th})\) is the cumulative distribution function (CDF) of the standard normal distribution.

$\implies$ Probability of being labeled as ``watermarked'' converges to a positive constant. Typical value $z_{th}=4$, $1-\Phi(z_{th})\approx 0.00003167$.

\item[S3.] If $q < \gamma$, then $\mathbb{E}[z]  \propto - \sqrt {T_e}$, and $T_c$ does not exist.

\noindent{\bf How?} A symmetric bound as equation \ref{eq:Cbound} is: for $y\leq T_e q$,
$$
\Pr (\|T\|_G \geq y) \leq \exp (-T_e D(\frac{k}{T_e}||q)) 
$$
Take $y=z_{th}\sqrt{T_e\gamma(1-\gamma)}+\gamma T_e \geq qT_e$,
\begin{align*}
    \Pr(z > z_{th}) \leq \Pr(z \geq z_{th})   \leq \exp\left(-T_eD(\frac{\sqrt{\gamma(1-\gamma)} z_{th}}{\sqrt{T_e}}+\gamma || q)\right)
\end{align*}

$\implies$ Probability of being labeled ``watermarked'' converges to 0 exponentially w.r.t. $T_e$.
\end{enumerate}

\subsection{\bf Case 2: Independent case} Assume the $c+1$-grams' colors are independent. 

\begin{enumerate}
\item[S1.] If $q > \gamma $, 
$\mathbb{E}[z]  \propto \sqrt {T_e}$ and  $T_c=\frac{\gamma(1-\gamma)z_{th}^2}{(q-\gamma)^2}$

\noindent{\bf How?} $\|T\|_G$ is a sum of $T_e$ different variables bounded by $[0,1]$. From the Hoeffding Inequality \citep{hoeffding1994probability}, for $t \geq 0$:
\begin{equation*}
\begin{split}
  \Pr (qT_e - \|T\|_G \geq t) &\leq \exp \left(-\frac{2t^2}{T_e} \right)
\end{split}
\label{eq:Hbound}
\end{equation*}
For $T_e\geq T_c$, take $t=(q-\gamma)\sqrt{T_e}(\sqrt{T_e}-\sqrt{T_c})\geq 0$, then:
\begin{equation*}
\begin{split}
\Pr(z \leq z_{th}) \leq &\exp \left(-2(q-\gamma)^2(\sqrt{T_e}-\sqrt{T_c})^2\right)
\end{split}
\end{equation*}

$\implies$ The probability of being labeled as ``not watermarked'' converges to 0 exponentially w.r.t. $T_e$.

\item[S2.] If $q=\gamma$, $\mathbb{E}[z] = 0$ and $T_c$ does not exist.

\item[S3.] If $q < \gamma$, $\mathbb{E}[z]  \propto - \sqrt {T_e}$ and $T_c$ does not exist. 

\noindent{\bf How?} A symmetric bound as from the Hoeffding Inequality \citep{hoeffding1994probability} gives, for $t \geq 0$:
\begin{equation}
\Pr ( \|T\|_G - qT_e \geq t) \leq \exp (-\frac{2t^2}{T_e} )
\label{eq:IndBound2}
\end{equation}
Take $t=(\gamma-q)(\sqrt{T_e}+\frac{\sqrt{\gamma(1-\gamma)}z_{th}}{\gamma-q})\sqrt{T_e} \geq 0$, and note that $\Pr(z > z_{th})\leq \Pr(z \geq z_{th})$:

\begin{equation*}
\begin{split}
        \Pr(z > z_{th})   \leq \exp(-2(\gamma-q)^2(\sqrt{T_e}+\frac{\sqrt{\gamma(1-\gamma)}z_{th}}{\gamma-q})^2) 
\end{split}
\end{equation*}

$\implies$ Still, the probability of being labeled as ``watermarked'' converges to 0 exponentially w.r.t. $T_e$.
\end{enumerate}

\newpage
\section{SCTS Efficacy Analysis}
\label{SCTS}

\subsection{Success Rate} 

For each $2$-grams' substitution attempt, the success probability
\begin{equation*}
\begin{split}
p_s &= \Pr(\text{$2-$gram is green, all $k$ candidates are green}) \\
&= \Pr(\text{$2-$gram is green}) \cdot \Pr(\text{one candidate is green})^k \\
&= p(1-\gamma^k)
\end{split}
\label{eq:psDefinition}
\end{equation*}

\subsection{Grouping} 
\label{Grouping}

For texts attacked by SCTS, $i.i.d.$ color distribution like assumption 2 does not hold anymore as one substitution may change the color of two tokens ($c=1$). We need to regroup so that every group's color is independent to reach similar bounds as theorem~\ref{theorem}. The grouping is as follows: 1. one $2$-gram as a group if it is preserved in SCTS, i.e., this token and the token preceding this token are not substituted); and 2. two adjacent $2$-grams as a group if they are not preserved because of the same substitution, i.e., the token being substituted and the token following it. We call the first type of groups ``old groups'' and the second type of groups ``new groups''. A corner case for one substitution only changes one $2$-gram as the word being substituted is the last (could not be the first for SCTS) is not considered. 

\subsection{Expected Green Ratio} 

From the assumption that candidate generation is independent of color, each group's number of green $2$-grams is independent, and $i.i.d.$ within old groups and new groups. Let the expectation of the ratio of green tokens for old groups be $p_o$ for notation. (``attempt fails" means that the attempt to substitute the first token in the $2-$gram fails in SCTS.)

\begin{align*}
p_o &= \Pr(\text{{$2-$gram is green}} \,|\, \text{{attempt fails}}) \\
&= \frac{\Pr(\text{{attempt fails}} \,|\, \text{{$2-$gram is green}}) \cdot \Pr(\text{{$2-$gram is green}})}{\Pr(\text{{attempt fails}})} \\
&= \frac{\gamma^k \cdot p}{1-p_s} \\
&= \frac{\gamma^k p}{1-p+\gamma^k p}
\end{align*}

Note the expectation of the ratio of green tokens for new groups is $\frac{\gamma}{2}$ and
$$p_o=\frac{\gamma^k p}{1-p+\gamma^k p} < \gamma \iff k>1+\log_\gamma \frac{1-p}{p(1-\gamma)}$$ 
Thus, equation~\ref{eq:qTeBound} holds as desired.

Let the expected number of green $2$-grams after SCTS attack be $\mathbb{E}_{T_e}=q(T_e)T_e$. Then we have:
\begin{equation*}
\mathbb{E}_{T_e}= p_s(\mathbb{E}_{T_e-2} + \gamma) + (1-p_s)(\mathbb{E}_{T_e-1} + p_o)
\end{equation*}
So
\begin{equation*}
T_e \cdot  q(T_e) = p_s(q(T_e-2)\cdot (T_e-2) + \gamma) + (1-p_s)(q(T_e-1) (T_e-1) + p_o)
\end{equation*}

Let $q = \lim_{T_e \to \infty} q(T_e)$,  $T_e \to \infty$, then

\begin{equation*}
T_e \cdot q = p_s(q \cdot (T_e-2) + \gamma) + (1-p_s)(q \cdot (T_e-1) + p_o)
\end{equation*}

\begin{equation*}
q = \frac{p_s \gamma + (1-p_s) p_o}{1 + p_s} = \gamma -\frac{\gamma(1-p\gamma^{k-1})}{1+(1-\gamma^k)p} < \gamma
\end{equation*}

\subsection{Success probability bound}

\begin{definition}[New $2$-gram ratio $r_n$] This is defined as the number of new (w.r.t the unattacked watermarked text) $2$-grams divided by $T_e$. Then $0\leq r_n\leq 1$.
\end{definition}

Based on the grouping strategy discussed in Appendix~\ref{Grouping}, we have $\frac{r_nT_e}{2}$ new groups and $(1-r_n)T_e$ old groups. Under the assumption of independence, the Hoeffding inequality gives: for \(t \geq 0\),
\begin{equation*}
\begin{split}
  \Pr (qT_e - \|T\|_G \geq t) \leq &\exp \left(-\frac{2t^2}{\frac{r_n T_e}{2} \cdot 2^2 + (1-r_n) T_e \cdot 1^2} \right) \\
  = &\exp \left(-\frac{2t^2}{(1+r_n)T_e} \right) \\
  \leq  &\exp \left(-\frac{t^2}{T_e} \right) 
\end{split}
\end{equation*}
So the only difference from equation~\ref{eq:IndBound2} is the denominator. When $q<\gamma$ for large enough $k$ and/or $T_e$, use similar techniques, i.e., Hoeffding's inequality, and have:
\begin{equation*}
\begin{split}
        \Pr(z > z_{th})   \leq \exp((\gamma-q(T_e))^2(\sqrt{T_e}+\frac{\sqrt{\gamma(1-\gamma)}z_{th}}{\gamma-q(T_e)})^2) 
\end{split}
\end{equation*}

Combining with equation~\ref{eq:limit_expression}, the probability of failing to evade watermarking exponentially converges to $0$ w.r.t. $T_e$.

\newpage
\section{SCTS LLM Calls}

\label{app1}

\begin{definition}[Number of LLM calls $N_{T_e}$] The number of LLM calls needed for SCTS on one sample with $T_e$ $2$-grams.
\end{definition}

\begin{equation}
\begin{split}
\mathbb{E} [N_{T_e}] \leq& \max \Bigg\{ \frac{1}{2} \left( \frac{1}{\gamma} - \left( \frac{1}{1-(1 - \gamma)^k} - 1 \right)k \right),  \frac{1-\gamma^k}{1-\gamma} \Bigg\}T_e \\
< &\max \left\{ \frac{1}{2\gamma}, \frac{1}{1-\gamma} \right\}T_e
\end{split}
\label{eq:LLMcallsbound}
\end{equation}

As \( T_e \to \infty \) and $k \to \infty $, the expected number of LLM calls per $c+1$-gram can be estimated. This is crucial for scalability.
\begin{equation}
    \lim_{T_e, k \to \infty} \frac{\mathbb{E}[N_{T_e}]}{T_e}  = \frac{p\gamma + (1-p)(1-\gamma)}{(1+p)\gamma(1-\gamma)}
    \label{eq:LLMcallslimit}
\end{equation}
For example, with \( p=0.5 \) and \( \gamma=0.25 \), this value is \( \frac{16}{9} \). Also, $N_{T_e}$ will not deviate far from its expectation with high probability.
 \begin{equation}
    \Pr (|\frac{N_{T_e}-\mathbb{E}[N_{T_e}]}{T_e}| \geq t) \leq \frac{C}{T_et^2}
    \label{SCTSPrBound}
\end{equation}
For $t>0$ and constant
\begin{equation*}
\begin{split}
C\coloneqq &\max \Bigg\{\frac{1}{2} \left(\frac{1 - \gamma}{\gamma^2}-\frac{k^2 (1 - \gamma)^k}{(1 - (1 - \gamma)^k)^2}\right), \frac{\gamma}{(1-\gamma)^2} - \frac{\gamma^k ((2k-1)(1-\gamma) + \gamma^k)}{(1 - \gamma)^2}
\Bigg\}\\
<& \max \Bigg\{  \frac{1-\gamma}{2\gamma^2}, \frac{\gamma}{(1-\gamma)^2}  \Bigg\}
\end{split}
\end{equation*}


\subsection{Derivation}
The grouping is the same as in Appendix \ref{Grouping}.

\begin{definition}[$N_{new}$] The number of LLM calls for a new \(2\)-grams group. \end{definition}
\begin{definition}[$N_{old}$] The number of LLM calls for a old \(2\)-grams group. \end{definition}
We have
\begin{equation*}
\Pr(N_{new}=i) = \frac{\Pr(G_{\gamma}=i)}{\Pr(G_{\gamma} \leq k)}, \quad i=1,2,\ldots,k
\end{equation*}
where \(G_{\gamma}\) is a geometric distribution with mean \(\frac{1}{\gamma}\). So,
\begin{equation*}
\begin{split}
\mathbb{E} [N_{new}] &= \mathbb{E} [G_\gamma | G_\gamma \leq k] \\
&= \frac{1}{\gamma} - \left(\frac{1}{1-(1 - \gamma)^k}-1\right)k \\
&< \frac{1}{\gamma}
\end{split}
\end{equation*}

For the old \(2\)-gram group, we have:
\begin{align*}
\Pr(N_{old}=i) &= \Pr(G_{1-\gamma}=i), \quad  1 \leq i\leq k-1 \nonumber\\
\Pr(N_{old}=k) &= \gamma^{k-1} \nonumber \\
&= \Pr(G_{1-\gamma}=k) + \Pr(G_{1-\gamma}>k)
\end{align*}
Therefore,
\begin{equation*}
\mathbb{E} [N_{old}] = \frac{1-\gamma^k}{1-\gamma}
< \frac{1}{1-\gamma}
\end{equation*}
Note that
\begin{equation*}
\begin{split}
\mathbb{E} [N_{T_e}] &= \left(r_n \frac{\mathbb{E} [N_{new}]}{2}  + (1-r_n)\mathbb{E} [N_{old}]\right)T_e \\
&\leq \max \left\{ \frac{\mathbb{E} [N_{new}]}{2}, \mathbb{E} [N_{old}] \right\} T_e
\end{split}
\end{equation*}

So equation \ref{eq:LLMcallsbound} holds.

\noindent{\bf Expectation limit (equation \ref{eq:LLMcallslimit}) derivation:} Denote $\mathbb{E} [N_{old}] = n_{o}$ and
$p_i^s= \Pr(\text{success at the $i$-th time}) = p \gamma^{i-1} (1-\gamma)$
for notational simplicity.


\begin{align*}
N_{T_e}= &\sum_{i=1}^{k} p_i^s (N_{T_e-2} + i) +(1 - p_s)(N_{T_e-1} + n_o)
\end{align*}
 Let \( T_e \to \infty \), $\lim_{T_e \to \infty} \frac{N_{T_e}}{T_e}=R$
\begin{align}
T_eR= &\sum_{i=1}^{k} p_i^s((T_e-2)R + i) \nonumber + (1-p_s)((T_e-1)R + n_o)
\end{align}

\[
R = \frac{\sum_{i=1}^{k} i p_i^s + (1-p_s)n_o}{1 + p_s}
\]
From this, we can get \( R \) for any \( k \), but the math is cumbersome. For simpler math, let \( k \to \infty \).
Then \( p_s = p \), \( N_{old}\) is a geometric distribution with mean \( n_o = \frac{1}{\gamma} \), 

so
\[
\sum_{i=1}^{\infty} i p_i^s = \frac{p}{1-\gamma}
\]
Thus, equation~\ref{eq:LLMcallslimit} holds.

\noindent{\bf Expectation bound by Chebyshev's inequality (equation \ref{SCTSPrBound} derivation):}

\begin{equation*}
\begin{split}
    Var(N_{new})&= \frac{1 - \gamma}{\gamma^2}-\frac{k^2 (1 - \gamma)^k}{(1 - (1 - \gamma)^k)^2} \\
    &< \frac{1 - \gamma}{\gamma^2}
\end{split}
\end{equation*}

\begin{equation*}
\begin{split}
Var(N_{old}) &= \frac{\gamma}{(1-\gamma)^2} - \frac{\gamma^k ((2k-1)(1-\gamma) + \gamma^k)}{(1 - \gamma)^2} \\
&< \frac{\gamma}{(1-\gamma)^2}
\end{split}
\end{equation*}
From independence, we have
\begin{equation*}
\begin{split}
&Var(N_{T_e}) \\
=& \left(r_n \frac{Var(N_{new})}{2}  + (1-r_n) Var(N_{old})\right)T_e \\
\leq& \max \left\{ \frac{Var (N_{new})}{2}, Var(N_{old}) \right\} T_e \\
=&CT_e 
\end{split}
\end{equation*}

From Chebyshev's inequality, for $\tau>0$,
\begin{equation*}
\begin{split}
    &\Pr (|N_{T_e}-\mathbb{E}[N_{T_e}]| \geq \tau \sqrt{CT_e}) \leq \frac{1}{\tau^2}\\
\end{split}
\end{equation*}
Namely
\begin{equation*}
\begin{split}
    &\Pr (|\frac{N_{T_e}-\mathbb{E}[N_{T_e}]}{T_e}| \geq \tau \sqrt{\frac{C}{T_e}}) \leq \frac{1}{\tau^2}
\end{split}
\end{equation*}
Take $t=\tau \sqrt{\frac{C}{T_e}}$, then equation \ref{SCTSPrBound} hold.

\newpage
\section{Additional Experimental Details}
\label{app:additional_exp}

\subsection{Dataset}
Following ~\citet{kirchenbauer2023watermark}, we use the training set of \texttt{C4 dataset’s RealNewsLike subset} \citep{raffel2020exploring}. 
For each sample, we first check if it is at least 500 tokens. If it is, we keep only the last 500 tokens. For this truncated part, the first 100 tokens are used as the prompt, and the remaining 400 tokens are used as ``human-written'' text for this prompt.
The (watermarked) text is generated by the LLM with watermarked decoding, but keeping other configurations' default as in the original implementation, truncated to at most 400 tokens. To avoid corner cases where the watermarked text is too short, we remove samples that are less than 20 words long. The number of samples is set to $50$ and $10$ for \texttt{vicuna-7b-v1.5-16k} and \texttt{Llama-2-7b-chat-hf} respectively, for reasons related to computational overheads. \\

\noindent{\bf SCT Experiment.} 
For SCT accuracy, we log the first $1000$ SCT results, including the SCTS result and ground truth in the same configuration as the main experiments (measuring AUROC, ASR, \# LLM calls, and Running time) but in another run.
This additional experiment is performed for UMD and \texttt{vicuna-7b-v1.5-16k} for simplicity.

\subsection{Budgets}
\label{Budgets}
We take budget $\{0.05,0.1,0.15,0.2,0.25,0.30,0.35,0.40,0.45,0.50\}$. For \textsc{Rp} and \textsc{Sico}, we additionally perform unconstrained attacks (budget$=1$). For \textsc{Scts}, to enforce different budgets, we yield the attacked sample when a budget is reached (one more substitution will be out of the budget) and continue to run until the maximum budget is reached or \textsc{Scts} saturates. When \textsc{Scts} saturates, the saturated attacked sample will be used as the output for higher budgets that are not reached. As a result, one run of \textsc{Scts} can give out attacked samples for all budgets, while only one budget for SICO.

\subsection{Metrics}

For evaluation, we consider the metrics listed below.

\begin{enumerate}
\item {\bf Area Under the Receiver Operating Characteristic (AUROC)}. Calculated based on the human-generated text and LLM-generated text. {\em The lower the score is, the more effective the attack is}, and 0.5 corresponds with the random guess.
\item {\bf Attack Success Rate (ASR).} It is the ratio of samples that are not successfully detected with the default threshold $z_{th}=4$. {\em Larger the value, the better the attack}.
\item {\bf \# LLM calls.} This is the number of LLM calls, including the calls to the paraphraser for \textsc{Rp} and \texttt{vicuna-7b-v1.5-16k / Llama-2-7b-chat-hf} calls for other methods. Calls of candidate generator \texttt{distilroberta-base} do not count as that model is much smaller and faster. {\em Smaller the value, the more efficient the attack.}
\item {\bf Running time.} This is used for comparing the speed of different approaches. Note that all experiments were conducted on one NVIDIA H100 80GB HBM3 GPU with Driver Version 535.54.03 and CUDA Version 12.2 on Ubuntu 20.04.6 LTS. We use Python 3.11.5 while Python 3.8.16 is used for \textsc{Sico} and \textsc{Rp}. {\em Smaller the value, the more efficient the attack.}
\item {\bf Semantic similarity.} We measure semantic similarity using embeddings generated by the \texttt{flan-t5-xxl} sentence encoder. We report the average cosine similarity for this particular metric.  {\em Higher the value, the more information is preserved by the attack.}
\item {\bf Accuracy.} This is the accuracy of the SCT test.  {\em Higher the better.}
\item {\bf Confusion matrix.} This is the confusion matrix associated with the SCT test, modeled as a three-class classification problem. {\em Closer to the diagonal is the better.}
\end{enumerate}

\subsection{SICO Details}
\subsubsection{Prompt}
For the SICO prompt with a budget of $0.5$, we modify this part of the prompt
\begin{quote}
    \texttt{Based on the description, rewrite this to P2 style:}
\end{quote}
to
\begin{quote}
    \texttt{Based on the description, rewrite this to P2 style, changing at most 50\% of the words to achieve the goal.}
\end{quote}

For different budgets less than $1$ (Unconstrained), the percentage of words that can be changed (\texttt{50\%}) is updated to reflect the allocated budget accordingly.
\subsubsection{Training}
For \texttt{Llama-2-7b-chat-hf}, we use the prompt from corresponding Unigram's training for reasons that they share the same training data and computational overheads.

\subsection{SCTS prompt for UMD Min-SelfHash $c=1$ and Unigram}
\label{prompt}
Because UMD Min-SelfHash $c=1$ and Unigram are essentially fixed green list agnostic to context, an adapted prompt example is as follows:

\begin{tcolorbox}
\footnotesize
\texttt{Choose two words (  includes, contains), and generate a long uniformly random string of those words separated by ";". Previous phrases should have no influence on future phrases: includes; contains; includes; contains; contains; includes; contains; contains; contains; contains; contains; contains; contains; contains; includes; contains; contains; includes; contains; contains;} 
\end{tcolorbox}


\newpage
\section{Full Experiment Results}

\label{full}

\begin{figure}[htbp]
\centering
\includegraphics[width=\linewidth]{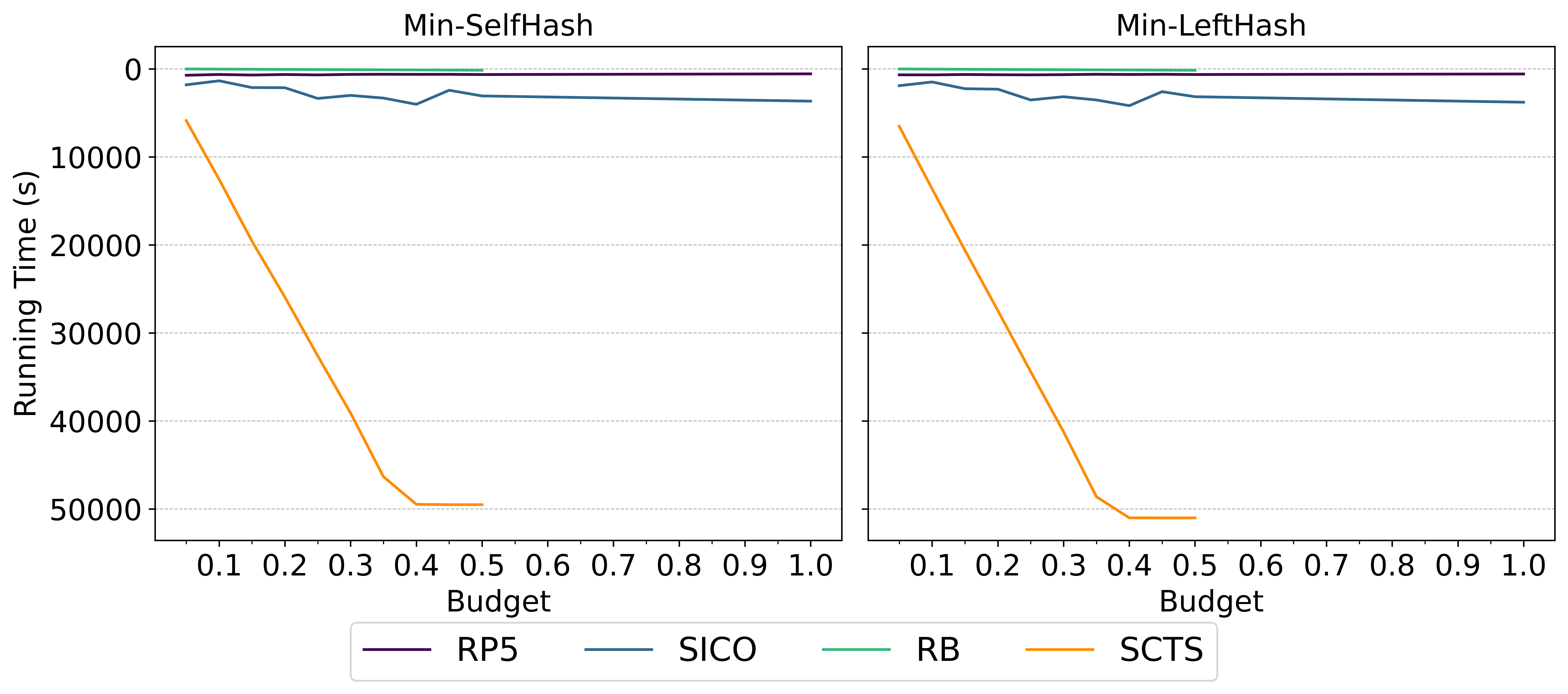}
\caption{Running time in seconds for vicuna-7b-v1.5-16k, 50 samples, UMD watermarking, $c=4$. A longer running time is needed for SCTS to perform a color-aware attack.}
\label{fig:R}
\end{figure}

\begin{figure}[htbp]
\centering
\includegraphics[width=\linewidth]{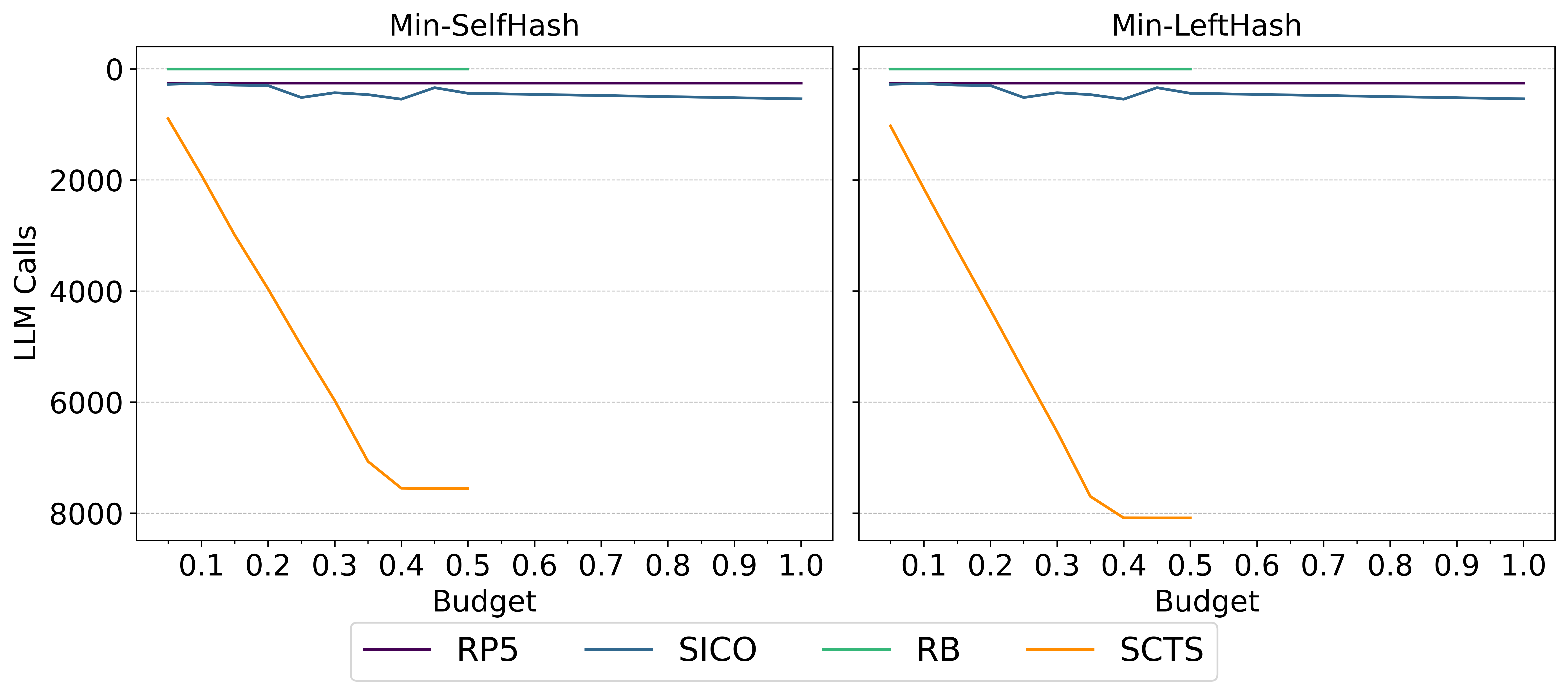}
\caption{\# LLM calls for vicuna-7b-v1.5-16k, 50 samples, UMD watermarking, $c=4$. The longer running time of SCTS mostly comes from more LLM calls for color information.}
\label{fig:L}
\end{figure}

\label{tables}
\begin{table*}[!t]
\centering
\caption{AUROC for \texttt{vicuna-7b-v1.5-16k}, 50 samples, UMD watermarking. SCTS achieves significantly lower AUROC under the same budget compared to other baselines and is the only method cross 0.5, and the trend is consistent over different $c$ and hashing methods.}
\label{tab:A1}
\begin{adjustbox}{width=\textwidth}
\begin{tabular}{@{}ccccccccccccccc@{}}
\toprule
c & Hashing & Method & 0      & 0.05   & 0.1    & 0.15   & 0.2    & 0.25   & 0.3    & 0.35   & 0.4    & 0.45   & 0.5    & 1 (Unconstrained)  \\
\midrule
        1 & left & RP1 & 0.9984 & 0.9984 & 0.9984 & 0.9984 & 0.9984 & 0.9984 & 0.9984 & 0.9984 & 0.9880 & 0.9984 & 0.9900 & 0.8036 \\ 
        1 & left & RP2 & 0.9984 & 0.9984 & 0.9984 & 0.9984 & 0.9984 & 0.9984 & 0.9964 & 0.9984 & 0.9880 & 0.9948 & 0.9872 & 0.9096 \\ 
        1 & left & RP3 & 0.9984 & 0.9984 & 0.9984 & 0.9984 & 0.9984 & 0.9964 & 0.9964 & 0.9984 & 0.9868 & 0.9948 & 0.9824 & 0.7428 \\ 
        1 & left & RP4 & 0.9984 & 0.9984 & 0.9984 & 0.9984 & 0.9984 & 0.9964 & 0.9964 & 0.9984 & 0.9868 & 0.9828 & 0.9828 & 0.8236 \\ 
        1 & left & RP5 & 0.9984 & 0.9984 & 0.9984 & 0.9984 & 0.9984 & 0.9964 & 0.9964 & 0.9984 & 0.9860 & 0.9804 & 0.9800 & 0.7136 \\ 
        1 & left & SICO & 0.9984 & 0.9960 & 0.9960 & 0.9960 & 0.9960 & 0.9960 & 0.9960 & 0.9956 & 0.9956 & 0.9852 & 0.9908 & 0.8404 \\ 
        1 & left & RB & 0.9984 & 0.9892 & 0.9816 & 0.9732 & 0.9600 & 0.9524 & 0.9320 & 0.9024 & 0.8744 & 0.8134 & 0.7488 & - \\
\bottomrule
\end{tabular}
\end{adjustbox}
\end{table*}

\begin{table*}[!t]
\centering
\caption{ASR for \texttt{vicuna-7b-v1.5-16k}, 50 samples, UMD watermarking, $z_{th}$=4. SCTS achieves significantly higher ASR under the same budget compared to other baselines, and the trend is consistent over different $c$ and hashing methods.}
\label{tab:D1}
\begin{adjustbox}{width=\textwidth}
{\small
\begin{tabular}{@{}ccccccccccccccc@{}}
\toprule
c & Hashing &Method & 0 (Unattacked) & 0.05 & 0.1 & 0.15 & 0.2 & 0.25 & 0.3 & 0.35 & 0.4 & 0.45 & 0.5 & 1 (Unconstrained) \\
\midrule
1 & left & RP1 & 0.18 & 0.18 & 0.18 & 0.18 & 0.18 & 0.18 & 0.18 & 0.18 & 0.18 & 0.20 & 0.28 & 0.96 \\ 
        1 & left & RP2 & 0.18 & 0.18 & 0.18 & 0.18 & 0.18 & 0.22 & 0.20 & 0.24 & 0.26 & 0.30 & 0.36 & 0.94 \\ 
        1 & left & RP3 & 0.18 & 0.18 & 0.18 & 0.18 & 0.18 & 0.24 & 0.26 & 0.28 & 0.32 & 0.36 & 0.46 & 1.00 \\ 
        1 & left & RP4 & 0.18 & 0.18 & 0.18 & 0.18 & 0.18 & 0.24 & 0.30 & 0.26 & 0.40 & 0.40 & 0.58 & 0.96 \\ 
        1 & left & RP5 & 0.18 & 0.18 & 0.18 & 0.20 & 0.18 & 0.24 & 0.30 & 0.28 & 0.44 & 0.42 & 0.62 & 1.00 \\ 
        1 & left & SICO & 0.18 & 0.28 & 0.28 & 0.28 & 0.30 & 0.30 & 0.32 & 0.30 & 0.28 & 0.40 & 0.40 & 0.82 \\ 
        1 & left & RB & 0.18 & 0.44 & 0.60 & 0.72 & 0.82 & 0.90 & 0.92 & 0.94 & 1.00 & 1.00 & 1.00 & - \\
\bottomrule
\end{tabular}
}
\end{adjustbox}
\end{table*}

\begin{table*}[!t]
\centering
\caption{Running time (s) for \texttt{vicuna-7b-v1.5-16k}, 50 samples, UMD watermarking. A longer running time is needed for SCTS to perform a color-aware attack, and the trend is consistent over different $c$ and hashing methods.}
\label{tab:R1}
\begin{adjustbox}{width=\textwidth}
{
\small
\begin{tabular}{@{}cccccccccccccc@{}}
\toprule
c & Hashing & Method & 0.05 & 0.1  & 0.15 & 0.2  & 0.25 & 0.3  & 0.35 & 0.4  & 0.45 & 0.5   & 1 (Unconstrained)  \\
\midrule
 1 & left & RP1 & 197 & 178 & 190 & 190 & 185 & 186 & 190 & 200 & 183 & 185 & 200 \\ 
        1 & left & RP2 & 374 & 363 & 376 & 375 & 367 & 374 & 381 & 387 & 373 & 361 & 357 \\ 
        1 & left & RP3 & 551 & 545 & 563 & 554 & 554 & 556 & 560 & 573 & 557 & 538 & 485 \\ 
        1 & left & RP4 & 734 & 730 & 757 & 744 & 729 & 732 & 736 & 747 & 732 & 707 & 595 \\ 
        1 & left & RP5 & 918 & 914 & 943 & 926 & 908 & 905 & 914 & 912 & 885 & 860 & 678 \\ 
        1 & left & SICO & 1926 & 1453 & 2195 & 2362 & 3522 & 3159 & 3438 & 4082 & 2568 & 3229 & 3758 \\ 
        1 & left & RB & 25 & 51 & 76 & 102 & 127 & 152 & 176 & 201 & 225 & 250 & - \\
\bottomrule
\end{tabular}
}
\end{adjustbox}
\end{table*}

\begin{table*}[!t]
\centering
\caption{\# LLM calls for \texttt{vicuna-7b-v1.5-16k}, 50 samples, UMD watermarking. The longer running time of SCTS mostly comes from more LLM calls for color information, and the trend is consistent over different $c$ and hashing methods.}
\label{tab:L1}
\begin{adjustbox}{width=\textwidth}
{
\scriptsize
\begin{tabular}{@{}cccccccccccccc@{}}
\toprule
c & Hashing & Method & 0.05 & 0.1  & 0.15 & 0.2  & 0.25 & 0.3  & 0.35 & 0.4  & 0.45 & 0.5   & 1 (Unconstrained)  \\
\midrule
1 & left    & RP1    & 50   & 50   & 50   & 50   & 50   & 50   & 50   & 50   & 50   & 50    & 50                 \\
1 & left    & RP2    & 100  & 100  & 100  & 100  & 100  & 100  & 100  & 100  & 100  & 100   & 100                \\
1 & left    & RP3    & 150  & 150  & 150  & 150  & 150  & 150  & 150  & 150  & 150  & 150   & 150                \\
1 & left    & RP4    & 200  & 200  & 200  & 200  & 200  & 200  & 200  & 200  & 200  & 200   & 200                \\
1 & left    & RP5    & 250  & 250  & 250  & 250  & 250  & 250  & 250  & 250  & 250  & 250   & 250                \\
1 & left    & SICO   & 275  & 263  & 291  & 299  & 514  & 428  & 462  & 544  & 337  & 438   & 538                \\
1 & left    & RB     & 0    & 0    & 0    & 0    & 0    & 0    & 0    & 0    & 0    & 0     & -                  \\
\bottomrule
\end{tabular}
}
\end{adjustbox}
\end{table*}

\begin{table*}[!t]
\centering
\caption{Semantic similarity for \texttt{vicuna-7b-v1.5-16k}, 50 samples, UMD watermarking. SCTS successfully preserves semantics, and the trend is consistent over different $c$ and hashing methods.}
\label{tab:S1}
\begin{adjustbox}{width=\textwidth}
{\small
\begin{tabular}{@{}cccccccccccccc@{}}
\toprule
c & Hashing & Method & 0.05 & 0.1  & 0.15 & 0.2  & 0.25 & 0.3  & 0.35 & 0.4  & 0.45 & 0.5   & 1 (Unconstrained)  \\
\midrule
        1 & left & RP1 & 1.0000 & 1.0000 & 1.0000 & 1.0000 & 1.0000 & 1.0000 & 1.0000 & 0.9987 & 0.9992 & 0.9928 & 0.8719 \\ 
        1 & left & RP2 & 1.0000 & 1.0000 & 0.9997 & 0.9987 & 0.9959 & 0.9964 & 0.9905 & 0.9908 & 0.9787 & 0.9780 & 0.8188 \\ 
        1 & left & RP3 & 1.0000 & 0.9998 & 0.9996 & 0.9981 & 0.9913 & 0.9878 & 0.9838 & 0.9803 & 0.9637 & 0.9559 & 0.7495 \\ 
        1 & left & RP4 & 1.0000 & 0.9998 & 0.9996 & 0.9976 & 0.9895 & 0.9811 & 0.9801 & 0.9674 & 0.9531 & 0.9355 & 0.6934 \\ 
        1 & left & RP5 & 1.0000 & 0.9998 & 0.9978 & 0.9973 & 0.9895 & 0.9809 & 0.9788 & 0.9645 & 0.9488 & 0.9198 & 0.6076 \\ 
        1 & left & SICO & 0.9989 & 1.0000 & 0.9906 & 0.9800 & 0.9847 & 0.9933 & 0.9958 & 0.9878 & 0.9674 & 0.9831 & 0.7065 \\ 
        1 & left & RB & 0.9890 & 0.9771 & 0.9639 & 0.9469 & 0.9337 & 0.9195 & 0.9054 & 0.8950 & 0.8863 & 0.8659 & - \\
\bottomrule
\end{tabular}
}
\end{adjustbox}
\end{table*}

\begin{table*}[!t]
\centering
\caption{AUROC for \texttt{vicuna-7b-v1.5-16k}, 50 samples, Unigram watermarking. We observe that Unigram is more robust than UMD, but still susceptible to SCTS with a consistent trend.}
\label{tab:A2}
\begin{adjustbox}{width=\textwidth}
\begin{tabular}{@{}ccccccccccccc@{}}
\toprule
Method & 0    & 0.05 & 0.1  & 0.15 & 0.2  & 0.25 & 0.3  & 0.35 & 0.4  & 0.45 & 0.5  & 1 (Unconstrained)  \\
\midrule
        RP1 & 0.9852 & 0.9852 & 0.9852 & 0.9852 & 0.9852 & 0.9856 & 0.9852 & 0.9852 & 0.9848 & 0.9844 & 0.9852 & 0.8852 \\ 
        RP2 & 0.9852 & 0.9852 & 0.9852 & 0.9852 & 0.9852 & 0.9856 & 0.9848 & 0.9844 & 0.9844 & 0.9840 & 0.9852 & 0.8708 \\ 
        RP3 & 0.9852 & 0.9852 & 0.9852 & 0.9852 & 0.9852 & 0.9856 & 0.9848 & 0.9820 & 0.9844 & 0.9840 & 0.9836 & 0.8700 \\ 
        RP4 & 0.9852 & 0.9860 & 0.9852 & 0.9852 & 0.9852 & 0.9856 & 0.9848 & 0.9840 & 0.9844 & 0.9840 & 0.9832 & 0.8564 \\ 
        RP5 & 0.9852 & 0.9860 & 0.9852 & 0.9852 & 0.9852 & 0.9856 & 0.9848 & 0.9840 & 0.9844 & 0.9832 & 0.9832 & 0.8088 \\ 
        SICO & 0.9852 & 0.9860 & 0.9860 & 0.9856 & 0.9860 & 0.9860 & 0.9824 & 0.9836 & 0.9848 & 0.9840 & 0.9488 & 0.8160 \\ 
        RB & 0.9852 & 0.9852 & 0.9856 & 0.9848 & 0.9848 & 0.9836 & 0.9736 & 0.9700 & 0.9668 & 0.9616 & 0.9584 & - \\ 
        SCTS & 0.9852 & 0.9820 & 0.9736 & 0.9552 & 0.9132 & 0.8488 & 0.7324 & 0.5932 & 0.4428 & 0.3288 & 0.2604 & - \\ 
\bottomrule
\end{tabular}
\end{adjustbox}
\end{table*}

\begin{table*}[!t]
\centering
\caption{ASR for \texttt{vicuna-7b-v1.5-16k}, 50 samples, Unigram watermarking, $z_{th}$=4. Unigram's unattacked ASR is higher, but still susceptible to SCTS with a consistent trend compared to UMD.}
\label{tab:D2}
\begin{adjustbox}{width=\textwidth}
\begin{tabular}{@{}ccccccccccccc@{}}
\toprule
Method & 0    & 0.05 & 0.1  & 0.15 & 0.2  & 0.25 & 0.3  & 0.35 & 0.4  & 0.45 & 0.5  & 1 (Unconstrained)  \\
\midrule
        RP1 & 0.62 & 0.62 & 0.62 & 0.62 & 0.62 & 0.62 & 0.62 & 0.62 & 0.64 & 0.64 & 0.64 & 1.00 \\ 
        RP2 & 0.62 & 0.62 & 0.62 & 0.62 & 0.62 & 0.62 & 0.66 & 0.62 & 0.66 & 0.64 & 0.70 & 0.92 \\ 
        RP3 & 0.62 & 0.62 & 0.62 & 0.62 & 0.62 & 0.66 & 0.68 & 0.66 & 0.68 & 0.66 & 0.72 & 1.00 \\ 
        RP4 & 0.62 & 0.62 & 0.62 & 0.62 & 0.62 & 0.64 & 0.68 & 0.66 & 0.72 & 0.74 & 0.78 & 0.96 \\ 
        RP5 & 0.62 & 0.62 & 0.62 & 0.62 & 0.62 & 0.64 & 0.68 & 0.68 & 0.76 & 0.74 & 0.78 & 0.98 \\ 
        SICO & 0.62 & 0.66 & 0.64 & 0.62 & 0.66 & 0.64 & 0.70 & 0.66 & 0.74 & 0.68 & 0.80 & 0.92 \\ 
        RB & 0.62 & 0.66 & 0.70 & 0.76 & 0.82 & 0.88 & 0.90 & 0.90 & 0.92 & 0.90 & 0.94 & - \\ 
        SCTS & 0.62 & 0.76 & 0.90 & 0.94 & 1.00 & 1.00 & 1.00 & 1.00 & 1.00 & 1.00 & 1.00 & - \\ 
\bottomrule
\end{tabular}
\end{adjustbox}
\end{table*}

\begin{table*}[!t]
\centering
\caption{Running time in seconds for \texttt{vicuna-7b-v1.5-16k}, 50 samples, Unigram watermarking. The trend is consistent from UMD, while the shorter running time is more due to implementation rather than methods per se.}
\label{tab:R2}
\begin{adjustbox}{width=\textwidth}
\begin{tabular}{@{}cccccccccccccc@{}}
\toprule
Method & 0.05 & 0.1  & 0.15 & 0.2  & 0.25 & 0.3  & 0.35 & 0.4  & 0.45 & 0.5  & 1 (Unconstrained)  \\
\midrule
        RP1 & 92 & 89 & 89 & 90 & 93 & 92 & 90 & 93 & 92 & 92 & 89 \\ 
        RP2 & 159 & 158 & 158 & 159 & 159 & 172 & 160 & 155 & 151 & 156 & 167 \\ 
        RP3 & 220 & 221 & 224 & 228 & 226 & 239 & 222 & 215 & 215 & 220 & 234 \\ 
        RP4 & 287 & 288 & 297 & 290 & 294 & 306 & 282 & 277 & 280 & 279 & 286 \\ 
        RP5 & 369 & 365 & 366 & 355 & 363 & 371 & 345 & 341 & 339 & 338 & 332 \\ 
        SICO & 1716 & 1240 & 2093 & 2123 & 3380 & 2930 & 3351 & 4066 & 2588 & 3108 & 3809 \\ 
        RB & 5 & 10 & 15 & 19 & 24 & 28 & 33 & 37 & 41 & 44 & - \\ 
        SCTS & 2125 & 4415 & 6553 & 8939 & 11267 & 13681 & 16042 & 18347 & 20580 & 22785 & - \\ 
\bottomrule
\end{tabular}
\end{adjustbox}
\end{table*}

\begin{table*}[!t]
\centering
\caption{\# LLM calls for \texttt{vicuna-7b-v1.5-16k}, 50 samples, Unigram watermarking. The trend is consistent with UMD.}
\label{tab:L2}
\begin{adjustbox}{width=\textwidth}
\begin{tabular}{@{}cccccccccccccc@{}}
\toprule
Method & 0.05 & 0.1  & 0.15 & 0.2  & 0.25 & 0.3  & 0.35 & 0.4  & 0.45 & 0.5  & 1 (Unconstrained)  \\
\midrule
RP1    & 50   & 50   & 50   & 50   & 50   & 50   & 50   & 50   & 50   & 50   & 50                 \\
RP2    & 100  & 100  & 100  & 100  & 100  & 100  & 100  & 100  & 100  & 100  & 100                \\
RP3    & 150  & 150  & 150  & 150  & 150  & 150  & 150  & 150  & 150  & 150  & 150                \\
RP4    & 200  & 200  & 200  & 200  & 200  & 200  & 200  & 200  & 200  & 200  & 200                \\
RP5    & 250  & 250  & 250  & 250  & 250  & 250  & 250  & 250  & 250  & 250  & 250                \\
SICO   & 275  & 263  & 291  & 299  & 514  & 428  & 462  & 544  & 337  & 438  & 538                \\
RB     & 0    & 0    & 0    & 0    & 0    & 0    & 0    & 0    & 0    & 0    & -                  \\
SCTS   & 576  & 1175 & 1730 & 2353 & 2984 & 3666 & 4318 & 4945 & 5547 & 6176 & -                  \\
\bottomrule
\end{tabular}
\end{adjustbox}
\end{table*}

\begin{table*}[!t]
\centering
\caption{Semantic similarity for \texttt{vicuna-7b-v1.5-16k}, 50 samples, Unigram watermarking. The trend is consistent with UMD with slightly lower values.}
\label{tab:S2}
\begin{adjustbox}{width=\textwidth}
\begin{tabular}{@{}cccccccccccccc@{}}
\toprule
Method & 0.05 & 0.1  & 0.15 & 0.2  & 0.25 & 0.3  & 0.35 & 0.4  & 0.45 & 0.5  & 1 (Unconstrained)  \\
\midrule
RP1 & 1.0000 & 1.0000 & 1.0000 & 1.0000 & 0.9997 & 1.0000 & 1.0000 & 0.9973 & 0.9976 & 0.9949 & 0.8735 \\ 
        RP2 & 1.0000 & 1.0000 & 0.9981 & 0.9986 & 0.9974 & 0.9924 & 0.9888 & 0.9884 & 0.9933 & 0.9833 & 0.7835 \\ 
        RP3 & 1.0000 & 1.0000 & 0.9967 & 0.9986 & 0.9953 & 0.9924 & 0.9838 & 0.9832 & 0.9903 & 0.9751 & 0.6923 \\ 
        RP4 & 0.9991 & 1.0000 & 0.9966 & 0.9971 & 0.9935 & 0.9914 & 0.9824 & 0.9744 & 0.9809 & 0.9727 & 0.5879 \\ 
        RP5 & 0.9991 & 1.0000 & 0.9966 & 0.9971 & 0.9927 & 0.9903 & 0.9821 & 0.9713 & 0.9769 & 0.9688 & 0.5247 \\ 
        SICO & 0.9980 & 1.0000 & 0.9959 & 0.9883 & 0.9744 & 0.9829 & 0.9841 & 0.9812 & 0.9920 & 0.9556 & 0.5702 \\ 
        RB & 0.9809 & 0.9664 & 0.9503 & 0.9343 & 0.9175 & 0.8973 & 0.8794 & 0.8603 & 0.8406 & 0.8151 & - \\ 
        SCTS & 0.9801 & 0.9666 & 0.9441 & 0.9241 & 0.9047 & 0.8836 & 0.8653 & 0.8434 & 0.8233 & 0.8015 & - \\ 
\bottomrule
\end{tabular}
\end{adjustbox}
\end{table*}

\begin{table*}[!t]
\centering
\caption{AUROC for \texttt{Llama-2-7b-chat-hf}, 10 samples, UMD watermarking. The trend is consistent compared with \texttt{vicuna-7b-v1.5-16k} over different $c$ and hashing methods.}
\label{tab:A3}
\begin{adjustbox}{width=\textwidth}
\begin{tabular}{@{}ccccccccccccccc@{}}
\toprule
c & Hashing & Method & 0      & 0.05   & 0.1    & 0.15   & 0.2    & 0.25   & 0.3    & 0.35   & 0.4    & 0.45   & 0.5    & 1 (Unconstrained)  \\
\midrule
 1 & left & RP1 & 1.0000 & 1.0000 & 1.0000 & 1.0000 & 1.0000 & 1.0000 & 1.0000 & 1.0000 & 1.0000 & 0.9900 & 1.0000 & 0.7700 \\ 
        1 & left & RP2 & 1.0000 & 1.0000 & 1.0000 & 1.0000 & 1.0000 & 1.0000 & 1.0000 & 1.0000 & 1.0000 & 1.0000 & 1.0000 & 0.8900 \\ 
        1 & left & RP3 & 1.0000 & 1.0000 & 1.0000 & 1.0000 & 1.0000 & 1.0000 & 1.0000 & 1.0000 & 1.0000 & 0.9700 & 1.0000 & 0.7800 \\ 
        1 & left & RP4 & 1.0000 & 1.0000 & 1.0000 & 1.0000 & 1.0000 & 1.0000 & 1.0000 & 1.0000 & 1.0000 & 0.9700 & 0.9900 & 0.7700 \\ 
        1 & left & RP5 & 1.0000 & 1.0000 & 1.0000 & 1.0000 & 1.0000 & 1.0000 & 1.0000 & 1.0000 & 1.0000 & 0.9700 & 0.9900 & 0.7100 \\ 
        1 & left & SICO & 1.0000 & 1.0000 & 1.0000 & 1.0000 & 1.0000 & 1.0000 & 1.0000 & 1.0000 & 1.0000 & 1.0000 & 1.0000 & 0.9300 \\ 
        1 & left & RB & 1.0000 & 1.0000 & 0.9700 & 0.9500 & 0.8500 & 0.8300 & 0.8100 & 0.8300 & 0.7900 & 0.7200 & 0.6100 & - \\
\bottomrule
\end{tabular}
\end{adjustbox}
\end{table*}

\begin{table*}[!t]
\centering
\caption{ASR for \texttt{Llama-2-7b-chat-hf}, 10 samples, UMD watermarking, $z_{th}$=4. The trend is consistent compared with \texttt{vicuna-7b-v1.5-16k} over different $c$ and hashing methods.}
\label{tab:D3}
\begin{adjustbox}{width=\textwidth}
\begin{tabular}{@{}ccccccccccccccc@{}}
\toprule
c & Hashing &Method & 0 (Unattacked) & 0.05 & 0.1 & 0.15 & 0.2 & 0.25 & 0.3 & 0.35 & 0.4 & 0.45 & 0.5 & 1 (Unconstrained) \\
\midrule
        1 & left & RP1 & 1.0000 & 1.0000 & 1.0000 & 1.0000 & 1.0000 & 1.0000 & 1.0000 & 1.0000 & 1.0000 & 0.9000 & 1.0000 & 0.0000 \\ 
        1 & left & RP2 & 1.0000 & 1.0000 & 1.0000 & 1.0000 & 1.0000 & 1.0000 & 1.0000 & 0.9000 & 0.9000 & 0.9000 & 0.9000 & 0.2000 \\ 
        1 & left & RP3 & 1.0000 & 1.0000 & 1.0000 & 1.0000 & 1.0000 & 1.0000 & 1.0000 & 0.9000 & 0.8000 & 0.9000 & 0.8000 & 0.0000 \\ 
        1 & left & RP4 & 1.0000 & 1.0000 & 1.0000 & 1.0000 & 1.0000 & 1.0000 & 0.9000 & 0.9000 & 0.8000 & 0.9000 & 0.6000 & 0.0000 \\ 
        1 & left & RP5 & 1.0000 & 1.0000 & 1.0000 & 1.0000 & 1.0000 & 1.0000 & 0.9000 & 0.9000 & 0.8000 & 0.8000 & 0.6000 & 0.0000 \\ 
        1 & left & SICO & 1.0000 & 0.8000 & 0.8000 & 0.8000 & 0.8000 & 0.8000 & 0.8000 & 0.8000 & 0.8000 & 0.8000 & 0.8000 & 0.3000 \\ 
        1 & left & RB & 1.0000 & 0.5000 & 0.4000 & 0.3000 & 0.2000 & 0.2000 & 0.1000 & 0.0000 & 0.0000 & 0.0000 & 0.0000 & - \\
\bottomrule
\end{tabular}
\end{adjustbox}
\end{table*}

\begin{table*}[!t]
\centering
\caption{Running time in seconds for \texttt{Llama-2-7b-chat-hf}, 10 samples, UMD watermarking. The trend is consistent compared with \texttt{vicuna-7b-v1.5-16k} over different $c$ and hashing methods.}
\label{tab:R3}
\begin{adjustbox}{width=\textwidth}
{\small
\begin{tabular}{@{}cccccccccccccc@{}}
\toprule
c & Hashing & Method & 0.05 & 0.1  & 0.15 & 0.2  & 0.25 & 0.3  & 0.35 & 0.4  & 0.45 & 0.5   & 1 (Unconstrained)  \\
\midrule
1 & left & RP1 & 34 & 43 & 35 & 34 & 36 & 36 & 36 & 36 & 37 & 34 & 35 \\ 
        1 & left & RP2 & 61 & 73 & 61 & 62 & 64 & 65 & 63 & 62 & 65 & 62 & 62 \\ 
        1 & left & RP3 & 92 & 105 & 87 & 91 & 94 & 92 & 91 & 90 & 91 & 91 & 83 \\ 
        1 & left & RP4 & 119 & 128 & 118 & 116 & 126 & 122 & 116 & 118 & 120 & 120 & 100 \\ 
        1 & left & RP5 & 148 & 156 & 148 & 142 & 153 & 150 & 144 & 142 & 149 & 144 & 110 \\ 
        1 & left & SICO & 1477 & 1082 & 1727 & 1809 & 3146 & 2726 & 3117 & 3733 & 2269 & 2904 & 3504 \\ 
        1 & left & RB & 5 & 11 & 16 & 21 & 26 & 32 & 37 & 42 & 47 & 52 & - \\
\bottomrule
\end{tabular}
}
\end{adjustbox}
\end{table*}

\begin{table*}[!t]
\centering
\caption{\# LLM calls for \texttt{Llama-2-7b-chat-hf}, 10 samples, UMD watermarking. The trend is consistent compared with \texttt{vicuna-7b-v1.5-16k} over different $c$ and hashing methods.}
\label{tab:L3}
\begin{adjustbox}{width=\textwidth}
{\scriptsize
\begin{tabular}{@{}cccccccccccccc@{}}
\toprule
c & Hashing & Method & 0.05 & 0.1  & 0.15 & 0.2  & 0.25 & 0.3  & 0.35 & 0.4  & 0.45 & 0.5   & 1 (Unconstrained)  \\
\midrule
1 & left & RP1 & 10 & 10 & 10 & 10 & 10 & 10 & 10 & 10 & 10 & 10 & 10 \\ 
        1 & left & RP2 & 20 & 20 & 20 & 20 & 20 & 20 & 20 & 20 & 20 & 20 & 20 \\ 
        1 & left & RP3 & 30 & 30 & 30 & 30 & 30 & 30 & 30 & 30 & 30 & 30 & 30 \\ 
        1 & left & RP4 & 40 & 40 & 40 & 40 & 40 & 40 & 40 & 40 & 40 & 40 & 40 \\ 
        1 & left & RP5 & 50 & 50 & 50 & 50 & 50 & 50 & 50 & 50 & 50 & 50 & 50 \\ 
        1 & left & SICO & 235 & 223 & 251 & 259 & 474 & 388 & 422 & 504 & 297 & 398 & 498 \\ 
        1 & left & RB & 0 & 0 & 0 & 0 & 0 & 0 & 0 & 0 & 0 & 0 & - \\
\bottomrule
\end{tabular}
}
\end{adjustbox}
\end{table*}

\begin{table*}[!t]
\centering
\caption{Semantic similarity for \texttt{Llama-2-7b-chat-hf}, 10 samples, UMD watermarking. The trend is consistent compared with \texttt{vicuna-7b-v1.5-16k} over different $c$ and hashing methods.}
\label{tab:S3}
\begin{adjustbox}{width=\textwidth}
{\small
\begin{tabular}{@{}cccccccccccccc@{}}
\toprule
c & Hashing & Method & 0.05 & 0.1  & 0.15 & 0.2  & 0.25 & 0.3  & 0.35 & 0.4  & 0.45 & 0.5   & 1 (Unconstrained)  \\
\midrule
        1 & left & RP1 & 1.0000 & 1.0000 & 1.0000 & 1.0000 & 1.0000 & 1.0000 & 1.0000 & 1.0000 & 0.9925 & 1.0000 & 0.8462 \\ 
        1 & left & RP2 & 1.0000 & 1.0000 & 1.0000 & 1.0000 & 1.0000 & 1.0000 & 0.9916 & 0.9887 & 0.9959 & 0.9896 & 0.7681 \\ 
        1 & left & RP3 & 1.0000 & 1.0000 & 1.0000 & 0.9954 & 1.0000 & 1.0000 & 0.9916 & 0.9790 & 0.9919 & 0.9789 & 0.6907 \\ 
        1 & left & RP4 & 1.0000 & 1.0000 & 1.0000 & 0.9954 & 0.9945 & 0.9983 & 0.9916 & 0.9671 & 0.9813 & 0.9523 & 0.6563 \\ 
        1 & left & RP5 & 1.0000 & 1.0000 & 1.0000 & 0.9954 & 0.9945 & 0.9983 & 0.9916 & 0.9671 & 0.9738 & 0.9303 & 0.4560 \\ 
        1 & left & SICO & 0.9988 & 0.9934 & 0.9960 & 0.9850 & 1.0000 & 1.0000 & 0.9932 & 1.0000 & 0.9696 & 0.9682 & 0.7528 \\ 
        1 & left & RB & 0.9894 & 0.9792 & 0.9660 & 0.9511 & 0.9415 & 0.9315 & 0.9148 & 0.8932 & 0.8994 & 0.8809 & - \\
\bottomrule
\end{tabular}
}
\end{adjustbox}
\end{table*}

\newpage

\end{document}